\documentclass{article}

\usepackage{arxiv}

\usepackage[utf8]{inputenc} 
\usepackage[T1]{fontenc}    
\usepackage{hyperref}       
\usepackage{url}            
\usepackage{amsfonts}       
\usepackage{nicefrac}       
\usepackage{microtype}      
\usepackage{graphicx}
\usepackage{natbib}

\usepackage{float}
\usepackage{shapepar}
\usepackage{listings}
\usepackage{amsthm}
\usepackage{mathtools}
\usepackage{graphics}
\usepackage{amssymb}
\usepackage{mathrsfs}
\usepackage{framed}
\usepackage{diagbox}
\usepackage{booktabs}
\usepackage{fancybox}
\usepackage{geometry}
\usepackage{multirow}
\usepackage{enumerate}
\usepackage{caption}
\usepackage{subcaption}
\usepackage{hyperref}
\usepackage{boldline}
\usepackage{vcell}
\usepackage{titlesec}
\usepackage{siunitx}
\usepackage[table]{xcolor}

\usepackage{makecell}
\usepackage{algorithmic}
\usepackage{lscape}
\usepackage{verbatim}
\usepackage{arydshln}

\usepackage[ruled,linesnumbered]{algorithm2e}
\newtheorem{theorem}{Theorem}[section]

\newtheorem{thm}{Theorem}[section]

\newcommand \R {\mathbb{R}}
\newcommand \C {\mathbb{C}}

\providecommand{\Jacobian}{\mathbf{J}}

\newcommand{\best}[1]{\cellcolor{black!12}#1}

\providecommand{\DeltaSigmaHat}{
\widehat{\delta \sigma}}

\providecommand{\BSigma}{\boldsymbol{\Sigma}}

\title{Deep Learning Based Reconstruction Methods for Electrical Impedance Tomography}

\date{} 					

\author{Alexander Denker  \\
Department of Computer Science\\ 
University College London
	\And
	Fabio Margotti \\ 
    Department of Mathematics \\ 
    Federal University of Santa Catarina
    \And 
    Jianfeng Ning \\ 
    School of Mathematics and Statistics \\ 
    Wuhan University
    \And 
    Kim Knudsen \\ 
    Department of Applied Mathematics and Computer Science \\ 
    Technical University of Denmark
    \And 
    Derick Nganyu Tanyu \\
    Centre for Industrial Mathematics (ZeTeM) \\
    University of Bremen
    \And 
    Bangti Jin \\ 
    Department of Mathematics \\ 
    The Chinese University of Hong Kong
    \And 
    Andreas Hauptmann \\
    Research Unit of Mathematical Sciences \\ 
    University of Oulu
    \And 
    Peter Maass \\
    Centre for Industrial Mathematics (ZeTeM) \\
    University of Bremen
}

\renewcommand{\shorttitle}{Deep Learning Based Reconstruction Methods for Electrical Impedance Tomography}

\hypersetup{
pdftitle={Deep Learning Based Reconstruction Methods for Electrical Impedance Tomography},
pdfsubject={},
pdfauthor={Alexander Denker},
pdfkeywords={First keyword, Second keyword, More},
}

\begin{document}
\maketitle

\begin{abstract}
Electrical Impedance Tomography (EIT) is a powerful imaging modality widely used in medical diagnostics, industrial monitoring, and environmental studies. The EIT inverse problem is about inferring the internal conductivity distribution of the concerned object from the voltage measurements taken on its boundary. This problem is severely ill-posed, and requires advanced computational approaches for accurate and reliable image reconstruction. Recent innovations in both model-based reconstruction and deep learning have driven significant progress in the field. In this review, we explore learned reconstruction methods that employ deep neural networks for solving the EIT inverse problem. The discussion focuses on the complete electrode model, one popular mathematical model for real-world applications of EIT. We compare a wide variety of learned approaches, including fully-learned, post-processing and learned iterative methods, with several conventional model-based reconstruction techniques, e.g., sparsity regularization, regularized Gauss-Newton iteration and level set method.  
The evaluation is based on three datasets: a simulated dataset of ellipses, an out-of-distribution simulated dataset, and the KIT4 dataset, including real-world measurements. Our results demonstrate that learned methods outperform model-based methods for in-distribution data but face challenges in generalization, where hybrid methods exhibit a good balance of accuracy and adaptability. 
\end{abstract}


\section{Introduction} 
Electrical impedance tomography (EIT) is a well-established imaging modality and is of
considerable practical interest in noninvasive
imaging and non-destructive testing, with several important medical applications, e.g., monitoring of lung functioning, detection of skin and breast cancer and location of epileptic foci \citep{Ho:04}. 
Further applications arise in geophysics to locate resistivity anomalies, e.g., minerals or contaminated sites \citep{ellis1994applied}. 
The  reconstruction problem is to infer the conductivity distribution inside the concerned object from boundary voltage measurements induced by several injected current patterns. 
The EIT inverse problem has been intensively studied, and there exists a very rich body of mathematical literature \citep{Borcea:2002,Uhlmann:2009}, since it was first formulated by \cite{Ca:80}.

This task is mathematically severely ill-posed, meaning that the reconstruction is highly susceptible to the presence of small noise in the data. To deal with the inherent ill-posedness, additional \textit{a priori} information has to be incorporated to allow for accurate reconstructions. Many \textit{model-based reconstruction methods} have been proposed based on variational regularization \citep{EnglHankeNeubauer:1996,ItoJinL2015}, e.g., Sobolev smoothness, total variation and sparsity.  These classical regularization functionals, like TV- or sparsity-based Tikhonov regularization, already exploit prior knowledge and promote the desired features, e.g., sparsity of edges or smoothness. Nonetheless, due to the severe ill-posedness and the high degree of non-linearity of the forward model in EIT, the resolution of the reconstructions obtained by these techniques has been modest at best.

The last years have witnessed significant progress in EIT reconstruction regarding both resolution and speed, and there is a fast growing body of literature on \textit{learned reconstruction methods} based on deep learning. This progress is primarily driven by recent innovations in deep learning, especially deep neural network architectures, efficient training algorithms (e.g., Adam \citep{KingmaBa:2015}), and powerful computing facilities, e.g., graphical processing units (GPUs).
Conceptually, learning the prior distribution with deep learning models from sufficiently large sets of (paired) training data (if available) potentially offers much greater flexibility than conventional hand-crafted priors, since the data manifold is often hard to capture by explicit mathematical models.

In this work, we investigate the use of deep learning models for EIT reconstruction of experimental data, and conduct a comprehensive comparative study of several representative approaches, including both traditional model-based methods and more recent learned reconstruction methods. There is a fast growing body of literature on deep learning-based approaches for EIT (including reviews, e.g., \cite{khan2019review}); see Section~\ref{sec:DL} for detailed discussions. However, most existing studies focus on showcasing the advantages of specific methods, and often compare their performance only with basic model-based methods. 

In the last years, several studies have provided comparative studies of deep learning techniques for different tomographic imaging modalities. For example, \cite{leuschner2021quantitative} and \cite{Kiss2025} have evaluated learned reconstruction methods for computed X-ray tomography (CT) using both simulated and real measured data, 
and have provided valuable insights into the performance of these methods on linear inverse problems. The work that is closest to our study is the recent work by \cite{Tanyu2025}, which focuses on deep learning-based EIT reconstruction. While building on similar themes, the present study involves key distinctions. First, we make use of the complete electrode model, which offers a more accurate and realistic representation of the forward process \citep{SomersaloCheney:1992}. Second, we investigate generalization properties of learned approaches by training models on simulated data and testing them on real measured data. This approach mirrors the notorious challenges of data-scarce applications, where large paired datasets are either unavailable or insufficient to train deep learning models accurately. Finally, \cite{Tanyu2025} does not include learned iterative methods \cite[Section 5.1.4]{arridge_acta} in the comparative study, which are often the most effective approach for linear inverse problems in terms of both reconstruction accuracy and generalization. It would be valuable to investigate how these methods perform in the context of nonlinear inverse problem of EIT.

The rest of the survey is organized as follows. In Section~\ref{sec:EIT}, we describe the complete electrode model for EIT and relevant mathematical and implementation issues. In Section \ref{sec:recon-alg}, we provide an overview of classical model-based reconstruction methods, which form the basis of learned approaches and also serve as benchmarks for the latter. In Section~\ref{sec:DL}, we summarize commonly used learned reconstruction approaches and their adaptations to EIT. In Section~\ref{sec:experimental_setup}, we provide details on the experimental setup, particularly the dataset used to train the learned reconstruction methods. Finally, in Section~\ref{sec:numerical_experiments}, we present numerical experiments conducted on both simulated and real data. We conclude the survey with further discussions in Section~\ref{sec:discussion}.

\section{Electrical impedance tomography}\label{sec:EIT}
Let $\Omega\subset \mathbb{R}^d$ ($d=2,3$) be an open bounded domain representing the concerned object with a Lipschitz boundary $\Gamma=\partial\Omega$.
In an EIT experiment, one applies an electrical current $j$ on
the boundary $\Gamma$ and then measures the resulting electrical
potential on the boundary $\Gamma$. The basic mathematical model, referred to as the continuum model, for the direct
problem is the following second-order elliptic partial differential equation (PDE)
\begin{equation}
\left\{\begin{aligned}
    -\nabla\cdot(\sigma\nabla u) &= 0,\quad \mbox{in } \Omega, \\ 
    \sigma\frac{\partial u}{\partial n} & = j, \quad \mbox{on }\Gamma, 
\end{aligned}\right.
\label{eqn:eit}
\end{equation}
where $n$ denotes the unit outward normal vector to the boundary $\Gamma$, and the conductivity $\sigma$ is assumed to belong to the admissible set 
\begin{equation}
    \mathcal{A}=\{\sigma\in L^\infty(\Omega): c_0\leq \sigma(x) \leq c_1 \mbox{ a.e. } \Omega\},
\end{equation}
where $c_0,c_1>0$ are given numbers. The Neumann boundary data $j$ should satisfy a compatibility condition $\int_{\Gamma} j {\rm d}s=0$. Under the normalization condition $\int_{\Gamma}u {\rm d}s=0$, problem \eqref{eqn:eit} has a unique solution $u$ in $\widetilde H^1(\Omega) = \{v\in H^1(\Omega): \int_{\Gamma}v {\rm d}s =0\}$. The Neumann to Dirichlet (NtD) operator $\Lambda_{\sigma,N}:j \mapsto \phi$ maps the Neumann data $j$ to the Dirichlet data $\phi=u|_{\Gamma}$ on $\Gamma$. 
The inverse problem in EIT aims at determining the
spatially-varying electrical conductivity $\sigma$ from the underlying NtD map $\Lambda_{\sigma,N}$. This problem was first formulated by \cite{Ca:80}, who also gave a uniqueness
result for the linearized problem. The mathematical theory of uniqueness
of the inverse problem with the complete NtD map $\Lambda_{\sigma,N}$ has received enormous attention (see, e.g.,
\cite{Is:06,Uhlmann:2009}). 
In practice, several input currents are injected, and then the induced electrical potentials
are measured (see, e.g., \cite{CheneyIsaacson:1992,jin2011function} for discussions on
optimal input currents), which contains some information about the NtD map $\Lambda_{\sigma,N}$. The inverse problem is to determine or at least to approximate the
unknown physical electrical conductivity $\sigma$ from 
partial knowledge of the NtD map.

Though mathematically well investigated, the continuum model \eqref{eqn:eit} does not capture two of the major complications when dealing with real measurements: (1) The input current is fed into the concerned object via an electrode with a certain spatial extension. For the input on each electrode, we can only have a total current injected on each electrode, and likewise we can measure the voltages on these electrodes. (2) There is a non-negligible contact impedance effect that models the transition of the current from the electrode to the surface of the object.
A more realistic mathematical model is the so-called complete electrode model (CEM) \citep{SomersaloCheney:1992}, which has been widely used in practical applications, e.g., industrial process tomography \citep{WilBec95}, geophysical exploration~\citep{Chaetal07}, medical diagnosis \citep{Boretal10} and non-destructive evaluation of concrete structures \citep{Karetal10}. Here, a set of $L$ electrodes is attached to the boundary $\Gamma$ of the object $\Omega$.
Then an input current is applied to some subset of these electrodes and the resulting potentials on all the electrodes are recorded, including the ones with applied current.
Mathematically, the model reads \allowdisplaybreaks
\begin{subequations}\label{eqn:cem}
\begin{align}
    -\nabla\cdot(\sigma\nabla u)&=0,\quad  \mbox{in }\Omega, \label{eqn:cem1} \\ 
u+z_\ell\sigma\frac{\partial u}{\partial n}&=U_\ell,\quad \mbox{on } e_l, \ell=1,2,\ldots,L,\label{eqn:cem2} \\ 
\int_{e_\ell}\sigma\frac{\partial u}{\partial n}{\rm d}s &=I_\ell, \quad \ell=1,2,\ldots, L, \label{eqn:cem3}\\
\sigma\frac{\partial u}{\partial n}&=0,\quad\mbox{on } \Gamma\backslash\cup_{\ell=1}^Le_\ell, \label{eqn:cem4}
\end{align}
\end{subequations}
where $\sigma\in\mathcal{A}$ is the electrical conductivity, $u\in H^1(\Omega)$ is the electric
potential, and $\{e_\ell\}_{\ell=1}^L\subset\Gamma$ denote the surfaces under the electrodes, which are assumed to be mutually disjoint. Further, $U_\ell\in\mathbb{R}$ and
$I_\ell\in\mathbb{R}$, $\ell=1,\ldots,L$, are the resulting electrode potentials and applied currents, respectively.

Equation \eqref{eqn:cem2} models the contact impedance effect: When the electrical currents are injected
into the object, a thin layer with high resistivity forms at the electrode-electrolyte interface, which causes potential drops at the electrodes as described by Ohm's law. 
Equations \eqref{eqn:cem3} and \eqref{eqn:cem4}  reflect the fact that the current injected through each electrode is completely confined to the electrode, and outside the electrodes there is no current. We denote the space of mean-free vectors by $\R_\diamond^L := \{ x \in \R^L : \sum_{\ell=1}^L x_\ell = 0\}$. The CEM imposes the conservation of charge (the compatibility condition for the current $I$), i.e., $I=(I_1,\ldots,I_L) \in \R_\diamond^L$, and the grounding condition $U =(U_1,\ldots,U_L)\in \R_\diamond^L$ for the electrode voltage $U$. 

The CEM \eqref{eqn:cem} is described for one single injection pattern $I \in \R_\diamond^L$. In practice, several different injection patterns are applied and the respective electrode voltages $U$ are measured. We denote the electrode voltages and charges for the $k$-th injection pattern by $U^{(k)}$ and $I^{(k)}$, for $k=1,\ldots,K$, with $K$ being the total number of injection patterns. 
Let $\mathbf{U}=(U^{(1)},\ldots, U^{(K)}) \in \R^{KL}$ and $\mathbf{I}=(I^{(1)},\ldots, I^{(K)}) \in \R^{KL}$ be the stacked vectors for all electrode voltages and current patterns. Let $F(\sigma)$ be the forward operator, i.e., the solution map of the CEM for a given $\sigma\in\mathcal{A}$, from the injection pattern $I$ to the electrical voltage $U$. Then the EIT direct problem reads
\begin{equation*}
   F(\sigma)I^{(k)}= U^{(k)},\quad k=1,\ldots, K,
\end{equation*}
or equivalently $F(\sigma)\mathbf{I}=\mathbf{U}$.

The goal of the EIT inverse problem in the context of the CEM \eqref{eqn:cem} is to reconstruct the conductivity $\sigma$ from the given electrode measurements $\mathbf{U}$. In practice, the measured voltages contain noise and below we denote the noisy voltage data by $\mathbf{U}^\delta$.

\subsection{Theoretical background}    
Now we review relevant theoretical results for the CEM \eqref{eqn:cem}. The weak formulation of problem \eqref{eqn:cem} reads: find $(u,U)\in\mathbb{H}$ such that
\begin{equation}\label{eqn:weak_form}
    \int_\Omega \sigma \nabla  u \cdot\nabla v{\rm d}x +\sum_{\ell=1}^L z_\ell^{-1}\int_{e_\ell} (u-U_\ell)(v-V_\ell){\rm d}s=\sum_{\ell=1}^L I_\ell V_\ell,\quad \forall (v,V)\in\mathbb{H},
\end{equation}
where the product space $\mathbb{H} := H^1(\Omega)\otimes \mathbb{R}^L_\diamond$ is equipped with the norm
\begin{equation}
   \Vert(u,U) \Vert_{\mathbb{H}}^2=\Vert u\Vert^2_{H^1(\Omega)}+\Vert U\Vert_{2}^2.
\end{equation}
\cite{SomersaloCheney:1992} showed that for any fixed $\sigma\in\mathcal{A}$, there exists a unique solution $(u,U)\in\mathbb{H}$ to problem \eqref{eqn:weak_form}, and it depends continuously on the current pattern $I$. Furthermore, the next theorem by \cite{jin2012analysis} gives an improved regularity of the solution. This result can be derived using the Neumann analogue of Meyers' gradient estimates \citep{Meyers:1963}. The improved regularity plays a crucial role in establishing the differentiability of the forward map.

\begin{theorem}[{\citealt[Theorem 3.3]{jin2012analysis}}]
Let $\lambda\in (0,1)$, and $\sigma(x)\in[\lambda,\lambda^{-1}]$ almost everywhere. Then there exists a constant $C_1(\Omega,\lambda,d)$, which depends only on $\Omega,d$ and $\lambda$, such that for any $q\in (2,C_1(\Omega,\lambda,d))$, there exists $C_2 =C_2(\Omega,d,\lambda,q)$ such that the solution $(u,U)\in \mathbb{H}$ to problem  \eqref{eqn:weak_form} satisfies the following estimate
    \begin{equation}
        \Vert u\Vert_{W^{1,q}(\Omega')}\le C_2 \Vert I\Vert_2.
    \end{equation}
\end{theorem}
Due to the change of the boundary conditions between the Neumann and Robin types, the state $u$ cannot achieve the full $H^2(\Omega)$ regularity (even if the conductivity $\sigma$ is smooth). Indeed, near the boundary $\partial e_\ell$ of the electrode $e_\ell$, the state $u$ exhibits weak singularities. We refer interested readers to \cite{DardeStaboulis:2016} for relevant discussions; see \cite{Grisvard:1985} for the general regularity theory. To address the limited regularity of solutions in the traditional CEM, \cite{hyvonen2017smoothened} proposed a smoothed version of the CEM. Further, \cite{Hyvonen:2004} showed that the CEM can be seen as a Galerkin approximation of the continuum model, and the corresponding current-to-voltage map can be seen as a discrete approximation of the traditional NtD operator $\Lambda_{\sigma,N}$. 

Due to the finite-dimensionality of the voltage data, the inverse problem in the CEM is inherently ill-posed; however, under certain conditions, uniqueness and stability can still be achieved. 
For the linearized reconstruction problem in the CEM, \cite{LechleiterRieder:2008} proved that for a piecewise polynomial conductivity on a fixed partition, a sufficiently large number of electrodes ensures the unique determination of the conductivity. More recently, \cite{harrach2019uniqueness} extended the results to the fully nonlinear case, and proved that the measurements from a sufficiently large number of electrodes uniquely determine the conductivity in any finite-dimensional subset of piecewise-analytic functions. \cite{harrach2019uniqueness} also established the Lipschitz stability for the inverse problem under the CEM. 

\subsection{FEM discretization}
\label{sec:fem}
Since the solution $(u,U)$ of the CEM \eqref{eqn:cem} is not available analytically, one often uses the finite element method (FEM) to obtain an approximate solution. We briefly recall the FEM discretization following the excellent description in \cite{LechleiterRider:2006}; see \cite{BrennerScott:2008} for the mathematical theory of the FEM. Specifically, one employs a shape-regular triangulation $\mathcal{T}$ of the domain $\Omega$ with $N$ nodes and defines a finite-dimensional subspace $V_h$ of the space $H^1(\Omega)$ to be the set of continuous piecewise linear functions subordinated to the triangulation $\mathcal{T}$. We denote by $\phi_j$ the standard hat function that takes the value 1 at the $j$th node and vanishes at all other nodes. Then any function $u_h\in V_h$ can be represented by
\begin{equation*}
    u_h(x) = \sum_{j=1}^N \alpha_j \phi_j(x), \quad \alpha_j\in \mathbb{R}.
\end{equation*}
Testing $(u_h,U)$ in the weak formulation \eqref{eqn:weak_form} against $(v,V)=(\phi_i,\mathbf{0})$ yields
\begin{equation}\label{eqn:CEM-fem1}
    \sum_{j=1}^N \alpha_j \int_\Omega \sigma\nabla \phi_j\cdot\nabla\phi_i {\rm d}x + \sum_{\ell=1}^L \frac{1}{z_\ell}\int_{e_\ell}\Big(\sum_{j=1}^N\alpha_j\phi_j-U_j\Big)\phi_i{\rm d}s=0,
\end{equation}
for $i=1, \dots, n$. Let $\mathbf{A}\in \mathbb{R}^{N\times N}$ be the admittance matrix and $\mathbf{B}\in \mathbb{R}^{N\times L}$ defined respectively by
\begin{equation*}
    A_{i,j} = \int_\Omega \sigma \nabla \phi_j\cdot\nabla\phi_i{\rm d}x + \sum_{\ell=1}^L\int_{e_\ell}\frac{1}{z_\ell}\phi_j\phi_i{\rm d}s \quad \mbox{and}\quad 
    B_{i,\ell} = -\frac{1}{z_\ell}\int_{e_\ell}\phi_i{\rm d}s.
\end{equation*}
Then with the vector $\boldsymbol\alpha=(\alpha_1,\ldots,\alpha_N)^\top\in\mathbb{R}^N$, equation \eqref{eqn:CEM-fem1} can be written as $$\mathbf{A}\boldsymbol{\alpha}+\mathbf{B}U =0.$$ Next we take into account the boundary conditions for $u_h$ arising from the CEM. Testing $(u_h,U)$ in the weak formulation \eqref{eqn:weak_form} against $(v,V)=(0,(\delta_{i,\ell})_{\ell=1}^L)$ (where $\delta_{i,\ell}$ denotes the Kronecker symbol), we obtain 
\begin{equation}
    \frac{1}{z_\ell}\int_{e_\ell}\Big(U_\ell-\sum_{j=1}^N\alpha_j\phi_j\Big){\rm d}s = I_\ell\quad \mbox{or} \quad
\label{eqn:CEM-fem2}
    U_\ell\frac{|e_\ell|}{z_\ell}-\sum_{j=1}^N\frac{\alpha_j}{z_\ell}\int_{e_\ell}\phi_j{\rm d}s = I_\ell,
\end{equation}
for $\ell=1,\ldots, L.$
Upon introducing the diagonal matrix $\mathbf{D}\in \mathbb{R}^{L\times L}$, with
\begin{equation*}
    D_{\ell,\ell} = \frac{1}{z_\ell}\int_{e_\ell}{\rm d}s = \frac{|e_\ell|}{z_\ell},
\end{equation*}
with $|e_\ell|$ being the Lebesgue measure of the $\ell$th electrode $e_\ell$, we may write \eqref{eqn:CEM-fem2} as $\mathbf{B}^\top \boldsymbol{\alpha} + \mathbf{D}U = I$. Finally, we obtain the linear system
\begin{equation}
    \begin{pmatrix}
        \mathbf{A}& \mathbf{B}\\
        \mathbf{B}^\top & \mathbf{D}
    \end{pmatrix}\begin{pmatrix}
        \boldsymbol{\alpha}\\ U
    \end{pmatrix} = \begin{pmatrix}
        0 \\ I
    \end{pmatrix}
\end{equation}
for obtaining the FEM solution $(u_h,U)$ of the forward problem \eqref{eqn:weak_form}. The system has to be augmented with the grounding condition $\sum_{\ell=1}^L U_\ell =0$. One can include this constraint by solving
\begin{equation}
    \begin{pmatrix}
        \mathbf{A}& \mathbf{B}\\
        \mathbf{B}^\top & \mathbf{D}\\
        0 & \mathbf{1}
    \end{pmatrix}\begin{pmatrix}
        \boldsymbol{\alpha}\\ U
    \end{pmatrix} = \begin{pmatrix}
        0 \\ I \\ 0
    \end{pmatrix},
\end{equation}
where $\mathbf{1}\in \mathbb{R}^{1\times L}$ is the row vector with all entries equal to 1. This approach is straightforward to implement but breaks the symmetry and positive definiteness property. Instead, one may consider a Lagrangian formulation for the constraint, which leads to a symmetric and positive definite system. Note that only the stiffness matrix $\mathbf{A}$ depends on the conductivity $\sigma$. That is, we only have to recompute $\mathbf{A}$ if the conductivity $\sigma$ changes.

The \textit{a priori} error analysis for the FEM solution $u_h$ is nontrivial since the solution $u$ of the CEM does not belong to $H^2(\Omega)$. This is due to the possible jumps of the conductivity $\sigma \in \mathcal{A}$ and because the Neumann boundary values do only belong to $H^s(\Gamma)$ for $s < 1/2$. In order to more effectively resolve the weak singularity of $u$, adaptive mesh refinement is a powerful strategy; see \cite{JinXuZou:2017,JinXu:2019} for related convergence analysis.

\section{Model-based reconstruction methods}\label{sec:recon-alg}
In this section we describe several model-based reconstruction methods for EIT. These methods roughly can be divided into two groups, i.e., reconstruction methods based on variational regularization (including classical Sobolev smoothness penalty, sparsity and level set method etc.) and direct methods (including D-bar method, factorization method and direct sampling method etc.).

\subsection{Variational reconstruction methods}
\label{sec:variational_methods}

Variational methods are widely employed to reconstruct solutions to ill-posed inverse problems. These methods are generally versatile and applicable to a broad class of inverse problems, and  are commonly referred to as Tikhonov regularization \citep{EnglHankeNeubauer:1996,ItoJinL2015}. The typical strategy is to formulate an optimization problem which minimizes a functional $J_\alpha$ comprising of a data discrepancy term $\psi$ and a suitable regularization term $\mathcal{R}$, suitably scaled by a regularization parameter $\alpha>0$. The regularization term $\mathcal{R}$ is used to overcome the inherent ill-posedness, by incorporating prior knowledge on the sought-for conductivity $\sigma$ into the functional which promotes physically meaningful reconstructions. Common choices for the regularizer $\mathcal{R}$ include the $L^2$ and $L^1$-norms, or the TV seminorm \citep{rudin1992nonlinear}. For example, the TV penalty is effective for recovering piecewise constant structures and capturing sharp discontinuities, see, e.g., \cite{Boretal10,MH22}. One popular choice of the Tikhonov functional is given by 
\begin{equation*}
    J_\alpha(\sigma) = \psi(\sigma)+ \alpha \mathcal{R}(\sigma),
\end{equation*}
with the regularization parameter $\alpha>0$ balancing the two terms, and the data fitting term $\psi(\sigma)$ given by 
\begin{equation*} 
\psi(\sigma) = \|F(\sigma)\mathbf{I}-\mathbf{U}^\delta\|_2^2 = \sum_{k=1}^K\|F(\sigma)I^{(k)} - U^{\delta,(k)}\|_2^2.
\end{equation*}
Then one takes a minimizer of the functional $J_\alpha(\sigma)$ over the admissible set $\mathcal{A}$ as an approximate solution to the EIT inverse problem. We discuss in detail the sparsity penalty in Section \ref{sec:sparsity} below. In practice, the problem has to be further discretized; see \citep{gehre2014analysis,FelisiRondi:2025} for convergence analysis of the FEM discretization. For the CEM, the forward map $F(\sigma)$ is nonlinear in $\sigma$, and the functional $J_\alpha$ is nonconvex and can have many local minima, thereby increasing the difficulty in minimizing the Tikhonov functional~$J_\alpha$.

To circumvent the nonconvexity, Newton-type methods are often used. These methods involve recursive linearization of the nonlinear operator $F(\sigma)\mathbf{I}$ around the current iterate $\sigma_k$. Let $F'(\sigma_k)\mathbf{I}$ be the Fr\'{e}chet derivative of $F(\sigma)\mathbf{I}$ at $\sigma_k$. Then the linearized model is given by
\begin{equation}\label{eqn:EIT-lin}
F(\sigma_k + \delta \sigma) \mathbf{I} \approx {F}(\sigma_k)\mathbf{I} + F'({\sigma_k})\mathbf{I}[\delta \sigma],
\end{equation}
around the current estimate $\sigma_k$. 
Then at iteration $k$, the fitting functional $\psi$ is approximated by a convex quadratic functional
\begin{equation}\label{LE}
  \psi_k(\delta \sigma ) = \|F'(\sigma_k)\mathbf{I}[\delta\sigma] + F(\sigma_k)\mathbf{I} - \mathbf{U}^\delta\|_2^2,  
\end{equation}
and the Tikhonov functional at the $k$th iteration reads
\begin{equation}\label{Tik}
J_{\alpha_k,k}(\delta\sigma) = \psi_k(\delta\sigma) + \alpha_k \mathcal{R}_k(\delta\sigma),
\end{equation}
where $\mathcal{R}_k$ is a penalization term at step $k$, possibly varying with $k$. Two commonly used choices are 
\begin{itemize}
    \item $\mathcal{R}_k(\delta\sigma) = \|\delta\sigma\|_{L^2(\Omega)}^2$ for the Levenberg--Marquardt method \citep{Levenberg:1944,Marquardt:1963,Hanke:1997};
    \item $\mathcal{R}_k(\delta\sigma) = \|\delta\sigma - (\sigma_k - \sigma_0)\|_{L^2(\Omega)}^2$ for the Iteratively Regularized Gauss--Newton (IRGN) method \citep{EnglHankeNeubauer:1996,MR2459012}.
\end{itemize}
It is a common practice to define $\psi_k$ using norms different from the original one in order to improve the reconstruction quality or to simplify optimization.
The Newton update is computed as 
$$\sigma_{k+1} = \sigma_k + \delta\sigma_k,$$ 
where $\delta\sigma_k$ minimizes the functional $J_{\alpha_k,k}$.

We employ a variation of the IRGN method, or the Gauss--Newton (GN) method. Specifically, we take the penalty $\mathcal{R}_k$ to be the total variation. In practice, for the discretized problem, we implement a smoothed version of the total variation defined by \citep{Boretal10} 
\begin{align*}
    \mathcal{R}_k(\sigma) = \sum_i \sqrt{({L}_i \sigma)^2 + \gamma},
\end{align*}
where ${L}_i$ denotes taking the discrete gradient at the $i$th node, and $\gamma > 0$ a smoothing parameter. Then the update $\delta \sigma_k$ can be approximated by (with one-step of reweighed least-squares) 
\begin{align}
    \label{eq:tv_update}
    \delta \sigma_k = (\Jacobian_{\sigma_k}^\top  \Jacobian_{\sigma_k} + \alpha \mathbf{L}^\top \mathbf{D}^{-1} \mathbf{L})^{-1}  \Jacobian_{\sigma_k}^\top\left(\mathbf{U}^\delta-F(\sigma_k)\mathbf{I} + \alpha \mathbf{L}^\top \mathbf{D}^{-1} \mathbf{L} \sigma_k \right),
\end{align}
with the diagonal matrix $\mathbf{D} = \text{diag}(\sqrt{(\mathbf{L}\sigma_k)^2 + \gamma})$ and $\mathbf{L}$ being the discrete finite difference matrix. Here we identify the function $\delta\sigma$ with the vector of its coefficients with respect to the given basis. In practice, computing the Jacobian matrix $\Jacobian_{\sigma_k}$ is typically quite expensive and depends on the choice of basis for $\sigma_k$. For piecewise constant basis functions, it requires solving $MK$ PDEs, where $M$ and $K$ denote the number of FEM elements (for a piecewise constant representation of $\sigma$) and current patterns, respectively. The method by \cite{Polydorides_2002} can reduce the computational burden, requiring the solution of only $K+L$ PDEs, where $L$ represents the number of electrodes; see e.g., \cite[Sec.~5.2.1]{margotti2015inexact}.

Iterative regularization methods for linear problems, e.g., gradient method and iterated--Tikhonov method, can be used in place of the Tikhonov method \eqref{Tik} to define an inner iteration that approximates a minimizer of the linearized problem \eqref{LE} with suitable accuracy. This idea forms the foundation of the inexact Newton method known as \texttt{REGINN} \citep{MR2570077,margotti2015inexact,MR1675352,M18}.
Alternatively, one can apply  gradient type methods directly to the functional $J_\alpha$ in order to get an approximate minimizer \cite[Sec. 11.1]{EnglHankeNeubauer:1996}. Then regularization may be achieved by prematurely terminating the iteration, using an appropriate stopping criterion, e.g., discrepancy principle \citep{EnglHankeNeubauer:1996,ItoJinL2015}.

One drawback of REGINN is its high computational cost, especially the computation of the Jacobian, which can make them slow for large-scale problems. To address this issue, one-step methods have been proposed, which employ a single iteration of a regularization scheme to compute an approximation. One classical example is the Linearized Reconstruction (LR) method, which performs one Newton step (at an initial guess $\sigma_0$). The LR method can also be interpreted as the first iteration of the LM or IRGN methods, with a suitable choice of the regularization functional $\mathcal{R}$. A well-known variant of the LR method is the NOSER algorithm~\citep{CheneyIsaacson:1990}, which was specifically developed for EIT and incorporates a tailored regularization strategy for the context.

In the numerical experiments in Section \ref{sec:numerical_experiments}, we employ the following version of the LR method for the discretized problem: we linearize the nonlinear forward operator  ${F}$ around a homogeneous background conductivity $\sigma_0$ and reconstruct a perturbation $\delta \sigma$ relative to $\sigma_0$ \citep{CheneyIsaacson:1990,dobson1994image,KaipioKolehmainenSomersalo:2000}. This results in the data-fidelity $\|\Jacobian_{\sigma_0}\delta\sigma - (\mathbf{U}^\delta-F(\sigma_0)\mathbf{I})\|_{\BSigma}^2$, under an additive Gaussian noise model $\mathcal{N}(\mathbf{0}, \BSigma)$. Then we discretize the perturbation $\delta \sigma \in \R^M$ using $M$ coefficients of a piecewise constant finite element expansion, and use the standard Tikhonov regularization, i.e., $\mathcal{R}(\delta\sigma)=\|\mathbf{R}\delta\sigma\|_2^2$, with $\mathbf{R} \in \R^{M \times M}$ being the regularizing matrix. Then the solution $\widehat{\delta\sigma}$ is given by 
\begin{align}
    \label{eq:lin_rec_one_step}
    \DeltaSigmaHat = (\Jacobian_{\sigma_0}^\top \BSigma^{-1} \Jacobian_{\sigma_0} + \alpha \mathbf{R}^\top \mathbf{R})^{-1} \Jacobian_{\sigma_0}^\top\BSigma^{-1} (\mathbf{U}^\delta-F(\sigma_0)\mathbf{I}).
\end{align}
Note that the matrix $(\Jacobian_{\sigma_0}^\top \BSigma^{-1} \Jacobian_{\sigma_0} +\alpha \mathbf{R}^\top \mathbf{R})^{-1}$ can be computed \textit{a priori}, leading to a computationally cheap reconstruction method, which is necessary for efficiently training post-processing methods (cf. Section~\ref{sec:lin_post}). Further, the Jacobian $\Jacobian_{\sigma_0}$ only has to be computed once, in contrast to the IRGN method. We use a smoothness prior, i.e., $\mathbf{R}^\top \mathbf{R} := \BSigma_\text{SM}^{-1}$ with $(\BSigma_\text{SM})_{i.j} = a \exp(- \| x_i - x_i \|_2^2/(2b)^2)$ with $x_i$ and $x_j$ being the positions of the mesh elements $i$ and $j$, with $a = 0.15^2$ and $b=0.2$.

\subsection{Sparsity reconstruction} 
\label{sec:sparsity}
The concept of sparsity is highly effective for modelling conductivity distributions that deviate from a known background $\sigma_0$, particularly when $\sigma$ consists of a largely homogeneous background with a few localized inclusions. Let $\delta\sigma^\dagger = \sigma^\dagger - \sigma_0$, where $\sigma^\dagger$ is the exact conductivity. Then we assume that $\delta\sigma$ admits a sparse expansion with respect to a chosen basis, frame, or dictionary $\{\psi_k\}$, meaning only a few nonzero coefficients. To enforce the sparsity of the reconstructed $\delta\sigma$, an $\ell^1$ penalty is employed \citep{DDD:2004}
\begin{equation}\label{eqn:psi} J_\alpha(\sigma) =  \psi(\sigma)+ \alpha \|\delta\sigma\|_{\ell^1},
\end{equation}
with 
$$\psi(\sigma)= \tfrac{1}{2}\|{F}(\sigma)\mathbf{I}-\mathbf{U}^\delta\Vert^2_2 \quad \mbox{and}\quad \|\delta\sigma\|_{\ell^1} = \sum_{k} |\langle\delta\sigma, \psi_k\rangle|.$$ 
Provided that the dictionary $\{\psi_k\}$ satisfies suitable regularity conditions, the minimization of $J_\alpha$ over the admissible set $\mathcal{A}$ is well-posed \citep{jin2012analysis}. The sparsity method was first developed by \cite{JinKhanMaass:2012} for the continuum model, and then was later adapted to the CEM by \cite{gehre2012sparsity} and we follow their derivation.

Optimization problems involving the $\ell^1$ penalty have attracted intensive interest in the last two decades \citep{BoneskyBrediesLorenzMaass:2007,BrediesLorenzMaass:2009,DDD:2004,WrightNowak:2009}. The primary computational challenges stem from the non-differentiable nature of the $\ell^1$ penalty coupled with the strongly nonlinearity of the term $\psi(\sigma)$. A basic scheme for iteratively updating $\delta \sigma_i$ and $\sigma_i = \sigma_0 + \delta \sigma_i$ to minimize $J_\alpha$ is formally given by
\begin{equation*}
\delta \sigma_{i+1} = \mathcal{S}_{s\alpha}(\delta \sigma_{i}-s {F}^\prime(\sigma_i)^{\ast}({F}(\sigma_i)\mathbf{I}-\mathbf{U}^\delta)),
\end{equation*}
where $s>0$ is the step size, ${F}^{\prime}(\sigma_i)$ denotes the G\^{a}teaux derivative of the map ${F}(\sigma)$ in $\sigma$ at $\sigma=\sigma_i$,  ${F}^\prime(\sigma_i)^{*}$ denotes the adjoint of  ${F}^{\prime }(\sigma_i)$, and $\mathcal{S}_\lambda(t)=\mbox{sign}(t)\max( |t|-\lambda,0)$ is the soft shrinkage operator.
However, the direct application of the algorithm often fails to produce satisfactory reconstructions. Instead, we implement the procedure detailed in Algorithm \ref{alg:sparse}, which includes two crucial steps: the computation of the gradient $\psi_s'$ (executed in Steps 4-5) and an appropriate step size selection (handled in Step 6).

\begin{algorithm}\small
\caption{Sparsity reconstruction for EIT.}
\label{alg:sparse}
\KwIn{ Background conductivity distribution $\sigma_0$, regularization parameter $\alpha$}
\KwResult{Reconstructed conductivity perturbation $\delta\sigma$}
Initialize perturbation estimate: $\delta\sigma_0 \leftarrow 0$\;
\For{i $\leftarrow$ 1, \ldots,}{
    Compute $\sigma_{i}=\sigma_0+\delta\sigma_{i}$\;
    Compute the gradient $\psi'(\sigma_{i})$\;
    Compute the $H_0^1$-gradient $\psi'_s(\sigma_{i})$\;
    Determine the step size $s_{i}$\;
    Update inhomogeneity by $\delta\sigma_{i+1} = \delta\sigma_{i} - s_{i}\psi_s'(\sigma_{i})$\;
    Threshold $\delta\sigma_{i+1}$ by $\mathcal{S}_{s_i\alpha}(\delta\sigma_{i+1})$\;
    Check stopping criterion.
}
\end{algorithm}

\noindent\textbf{Gradient evaluation} To adapt the basic algorithm to the CEM setting, the gradient $\psi'(\sigma)$ of the discrepancy term $\psi(\sigma)$ must be computed. The gradient can be decomposed into the individual applied current patterns:
\begin{align}
    {F}^\prime(\sigma_i)^{\ast}({F}(\sigma_i)\mathbf{I}-\mathbf{U}^\delta) = \sum_{k=1}^K F^\prime(\sigma_i)^{\ast}(F(\sigma_i)I^{(k)}-U^{\delta, (k)}).
\end{align}
For each current pattern $I^{(k)}$, the respective gradient can be efficiently  calculated via the adjoint technique. For notational simplicity we drop the superscript $(k)$ below. The adjoint problem reads: Find $(p,P) \in \mathbb{H}$ such that 
\begin{equation}
    \begin{cases}
        \begin{aligned}
            -\nabla \cdot(\sigma\nabla p) &= 0, && \text{in }\Omega, \\
            p + z_\ell \sigma \frac{\partial p}{\partial n} &= P_\ell, && \text{on }e_\ell, \ell=1,2,...,L, \\
            \int_{e_\ell} \sigma \frac{\partial p}{\partial n} ds &= U_\ell(\sigma) -U_\ell^\delta, && \text{for }\ell=1,2,...,L, \\
            \sigma \frac{\partial p}{\partial n} &= 0, && \text{on }\Gamma\backslash\cup_{\ell=1}^Le_\ell,
        \end{aligned}
    \end{cases}
\end{equation}
where ${U}(\sigma)=F(\sigma)I\in \mathbb{R}^L$ and $U^\delta$ is the electrode voltage measurements. Given the adjoint state $p$, the gradient $\psi'(\sigma)$ is given by 
\begin{equation*}
\psi'(\sigma)=-\nabla u(\sigma)\cdot \nabla p(\sigma).
\end{equation*}

Since $\psi'(\sigma)$ is defined by the duality pairing $\psi'(\sigma)[\lambda]=\langle \psi'(\sigma),\lambda\rangle_{L^2(\Omega)}$, it belongs to the dual space $(L^\infty(\Omega))'$ and may lack sufficient Sobolev regularity. To overcome this issue, we reformulate it using the $H_0^1(\Omega)$ inner product $\langle\cdot,\cdot\rangle_{H_0^1(\Omega)}$ by defining the Sobolev gradient $\psi'_s(\sigma)$ such that $\psi'(\sigma)[\lambda]=\langle \psi_s'(\sigma),\lambda \rangle_{H_0^1(\Omega)}$. Integration by parts leads to the following elliptic boundary value problem
\begin{equation}\label{eqn:grad-Sobolev}
\left\{\begin{aligned}-\Delta \psi_s'(\sigma)+\psi_s'(\sigma)&=\psi'(\sigma),\quad &&\mbox{in }\Omega,\\
\psi_s'(\sigma) &= 0,\quad &&\mbox{on } \Gamma.
\end{aligned}\right.
\end{equation}
The boundary condition explicitly encodes the assumption that the inclusions are in the interior of the domain $\Omega$ and $\sigma$ is constant on the boundary. The Sobolev gradient \citep{Neuberger:1997} is a regularized version of the $L^2(\Omega)$ gradient, and it defines a smoother admissible set $\mathcal{A}$ using the $H_0^1(\Omega)$ topology. Numerically, $\psi_s'(\sigma)$ is obtained by solving the Poisson equation \eqref{eqn:grad-Sobolev}. Using $\psi_s'$, we approximate $J_\alpha$ locally as
\begin{align*}
&J_\alpha(\sigma_0 + \delta \sigma) - J_\alpha(\sigma_0 + \delta \sigma_i) \\
\approx &
\langle\delta\sigma-\delta\sigma_i,\psi_s'(\sigma_i)\rangle_{H^1(\Omega)}
+\tfrac{1}{2s_i}\|\delta\sigma-\delta\sigma_i\|_{H^1(\Omega)}^2+\alpha\|\delta\sigma\|_{\ell^1},
\end{align*}
which is equivalent to
\begin{equation}\label{eqn:proxy}
\tfrac{1}{2s_i}\|\delta\sigma-(\delta\sigma_i-s_i\psi_s'(\sigma_i))
\|_{H^1(\Omega)}^2+\alpha\|\delta\sigma\|_{\ell^1}.
\end{equation}
The minimizer $\delta\sigma_{i+1}$ is given explicitly by
\begin{equation*}
\delta\sigma_{i+1}=\mathcal{S}_{s_i\alpha}(\delta\sigma_i-s_i\psi_s'(\sigma_i)).
\end{equation*}
This step zeros out small coefficients, and promotes the sparsity of $\delta\sigma$ with respect to the chosen basis $\{ \psi_k \}$.

\noindent \textbf{Step size selection} Classical gradient-based methods, e.g., steepest descent, often suffer from slow convergence. To improve the computational efficiency, the Barzilai-Borwein method \citep{BarzilaiBorwein:1988} proposes to approximate the Hessian by a scalar multiple of the identity matrix in a least-squares sense
\begin{equation*}
s_i=\arg\min_s\|s(\delta\sigma_i-\delta\sigma_{i-1})-(
\psi_s'(\sigma_i)-\psi_s'(\sigma_{i-1}))\|_{H^1(\Omega)}^2. 
\end{equation*}
This gives rise to one popular Barzilai-Borwein rule \citep{DaiHagerZhang:2006}
\begin{align*}
    s_i=\langle\delta\sigma_i-\delta\sigma_{i-1},\psi_s'
(\sigma_i)-\psi_s'(\sigma_{i-1})\rangle_{H^1(\Omega)}/\|\delta\sigma_i
-\delta\sigma_{i-1}\|_{H^1(\Omega)}^2.
\end{align*}
In practice, following \cite{WrightNowak:2009}, we enforce a weak monotonicity condition
\begin{align} \label{eqn:weak-mono}
&J_\alpha(\sigma_0+\mathcal{S}_{s\alpha}(\delta\sigma_i-s\psi_s'(\sigma_i)))\\
 \leq &
\max_{i-M+1\leq k\leq i}J_\alpha(\sigma_k)-\tau\frac{s}{2}
\|\mathcal{S}_{s\alpha}(\delta\sigma_i-s\psi'_s(\sigma_i))-\delta\sigma_i\|_{H^1(\Omega)}^2,\nonumber
\end{align}
where $\tau$ is a small positive constant and $M \geq 1$ is an integer. The step size $s$ is initialized using the above rule and then reduced geometrically by a factor $q\in(0,1)$ until the weak monotonicity condition \eqref{eqn:weak-mono} is satisfied. The iteration is terminated when $s_i$ falls below a threshold $s_\mathrm{stop}$ or after reaching the maximum number of iterations. In practice, one employs piecewise linear elements for approximating the state $u$ as well as the conductivity $\sigma$ in the sparsre recovery, whereas in all other methods we employ piecewise constant basis for approximating $\sigma$.

\subsection{Level set method}
\label{sec:level_set}
The level set method due to \cite{osher1988} paramaterizes a piecewise constant function using level sets of an auxiliary function. It is well-suited for EIT as we often deal with piecewise constant conductivities. The inverse problem is then cast as a fully nonlinear variational problem for the auxiliary function.  
The level set method was adapted to inverse problems by \cite{santosa1995} and in the  context of EIT with TV regularization~\citep{chung2005}. The original level set method was generalized in \cite{chan2004} to allow for multiple constant-value coefficients. It was successfully implemented for EIT~\citep{Amal} for the Kuopio Tomography Challenge \citep{Mikko}. We follow \cite{Amal,chan2004} for the description.

Let $\Omega \subset \R^2$ be a smooth bounded domain, and the strictly positive ${\sigma \in L^\infty(\Omega)}$ takes on four different 
conductive values $\sigma_i,\; i=1,2,3,4.$ Let $\Omega_i \subset \Omega$ denote the region defined by $\sigma(x) = \sigma_i.$ The level set parametrization uses two smooth auxiliary functions $\phi_1$ and $\phi_2$ defined on $\Omega$ to represent the regions
\begin{equation}\label{eq:regions}
    \begin{aligned}
        \Omega_1& = \{x\in \Omega: \phi_1>0, \phi_2>0\}, \quad
        \Omega_2 = \{x\in\Omega: \phi_1>0,\phi_2\leq 0\},\\
        \Omega_3 &= \{x\in \Omega: \phi_1\leq 0,\phi_2> 0\},\quad
        \Omega_4 = \{x\in \Omega: \phi_1\leq 0, \phi_2\leq 0\}.
    \end{aligned}
\end{equation}
The conductivity $\sigma$ is parametrized by the Heaviside function $H$ and $\phi_1,\phi_2$ as
\begin{align} \label{eq:compute_sigma}
        {\sigma}(\phi_1,\phi_2) =& \sigma_1 H(\phi_1)H(\phi_2)+\sigma_2 H(\phi_1)(1-H(\phi_2))
    \nonumber\\
    &+\sigma_3(1-H(\phi_1))H(\phi_2) +\sigma_4(1-H(\phi_1))(1-H(\phi_2)),
\end{align}
The approach allows four different conductivity values, but in the experiment we only segregate the domain $\Omega$ into three regions, i.e., the background, higher and lower conductivity. The specific values of $\sigma_i$ are fixed, e.g., estimated by prior knowledge, and hence the method is more qualitative than quantitative.

In the level set method, we cast the reconstruction task via the nonlinear functional
\begin{equation} \label{levelset_functional}
J({\phi}_1,{\phi}_2) =\!\tfrac{1}{2}\|(F(\sigma(\phi_1,\phi_2))-F(\sigma_0))\mathbf{I}-(\mathbf{U}^\delta-\mathbf{U}_0^\delta)\|_2^2 + \alpha |{\sigma}({\phi}_1,{\phi}_2)|_{\mathrm{TV}},
\end{equation}
where the regularization parameter $\alpha>0$ can be chosen using the validation data \citep{chan2004}, $|\cdot|_{\rm TV}$ denotes the total variation seminorm, and $\sigma_0$ is the homogeneous background conductivity. We use the full nonlinear forward model ${F}$ rather than a linearization, and the difference data rather than absolute data. To minimize the objective $J(\phi_1,\phi_2)$, we employ a gradient-descent method. The chain rule yields
\begin{equation*}
    \frac{\partial F}{\partial \phi_i} = {F}'(\sigma) \frac{\partial \sigma}{\partial \phi_i}, \quad i=1,2,
\end{equation*}
where ${ F}'(\sigma)$ is defined above, and formally
\begin{equation*}
    \begin{aligned}
        \frac{\partial \sigma}{\partial \phi_1} &= \delta(\phi_1)[\sigma_1 H(\phi_2)+\sigma_2(1-H(\phi_2))-\sigma_3H(\phi_2)-\sigma_4(1-H(\phi_2))],\\
    \frac{\partial \sigma}{\partial \phi_2} &= \delta(\phi_2)[\sigma_1H(\phi_1)-\sigma_2H(\phi_1)+\sigma_3(1-H(\phi_1))-\sigma_4(1-H(\phi_1))],
    \end{aligned}
\end{equation*}
where $\delta$ is the Dirac delta distribution. In order to avoid dealing with such singular functions numerically, we use smooth approximations of these two functions:
\begin{align*}
    \delta_\varepsilon(s)=\frac{\varepsilon}{\pi(s^2+\varepsilon^2)}\quad \mbox{and}\quad
    H_\varepsilon(s) &= \frac{1}{\pi}\tan^{-1}\left(\frac{s}{\varepsilon}\right)+\frac12,
\end{align*}
where the small parameter $\varepsilon>0$ controls the smoothness; in practice we choose $\varepsilon$ according to the mesh size $h$. The update in the $j+1$'th iteration is then given by
\begin{align*}
    {\phi}_{i}^{(j+1)}\ = {\phi}_{i}^{(j)} - \tau \frac{\mathrm{d}J}{\mathrm{d}{\phi}_{i}^{(j)}},\quad i=1,2,
\end{align*}
for a step length $\tau>0.$ 
The computational details including parameter choices follow \cite{Amal}. Note that the implementation uses re-initialization of the level set functions $\phi_i$ in order to keep them as signed distance functions. Moreover, the TV-functional is approximated smoothly and implemented via lagged diffusivity \citep{vogel1996,adesokan2019}. Further, a good initial guess for the level set method is required to reduce the risk of trapping into local minima. For the results below we use a segmented reconstruction based on a basic linearized method like \eqref{eqn:EIT-lin}.

For the numerical experiments below, we choose $\sigma_2 = 5$ and $\sigma_3 = 0.1$. These choices are somewhat arbitrary. In the functional $J(\phi_1,\phi_2)$, $\alpha = 5\mathrm{e}{-8}$ is fixed for all experiments. Moreover, for all reconstructions, the algorithm terminates after $1000$ iterations. Our implementation of the level set method uses the numerical algorithm by \cite{Amal}. The code was developed for 32-electrode data, so the 16-electrode data used in this study require a preprocessing step. In this adaption we introduce a minor modeling error that we do not handle explicitly. 

\subsection{Direct reconstruction methods}
In this part, we describe several direct methods for EIT imaging, which are specifically developed for concrete inverse problems, and essentially exploit the structure of the direct problems. The common advantage of these methods is that they are computationally efficient, and do not require the repeated application of the forward operator. However, unfortunately they also directly inherit the exponential instability to noise, hence need stabilization, generally provide reconstructions that tend to be overly smooth, and moreover, it is often challenging to include further prior knowledge into the reconstruction in the form of additional regularization. Nevertheless, direct methods can provide a proven convergent regularization, as it is the case for the D-bar algorithm \citep{KnudsenLassasSiltanen:2009}. Below we describe three direct methods, i.e., D-bar method \citep{SiltanenMuellerIsaacson:2000}, factorization method~\citep{HB2003} and direct sampling method \citep{chow2014direct}. There are other direct reconstruction methods, e.g., classic Calder\'on's method \citep{BikowskiMuller:2008,ShinMueller:2020}, 
enclosure method \citep{ikehata1998reconstruction} (with its recent learned variation \citep{sippola2025learned}), and the monotonicity method \citep{HarrachUllrich:2013}.

\subsubsection{D-bar method} 
 \label{sec:Dbar}
The D-bar method as implemented by \cite{SiltanenMuellerIsaacson:2000} relies on a reformulation of the conductivity equation into a Schr\"odinger-type equation and uses tools from complex scattering theory. It is based on the theoretical uniqueness proof and reconstruction algorithm by \cite{Nachman:1996}; see also later developments \citep{knudsenTamasan,AstalaPaivrinta:2006} and the three-dimensional counterpart \citep{Nachman1988,Novikov:1988,delbary2014a,KnudsenRasmussen2022}. The method was adapted to piecewise constant conductivities by \cite{KnudsenLassasMuellerSiltanen2007}.

The algorithm relies on the knowledge of the full Dirichlet-to-Neumann (DtN) map $\Lambda_{\sigma,D}$ for the continuum model~\eqref{eqn:eit}, which can be obtained as the inverse of the measured NtD map $\Lambda_{\sigma,N}$. In order to apply the algorithm to the CEM in~\eqref{eqn:cem}, we need to transform the measurements into a representation of the DtN map $\Lambda_{\sigma,D}$. 
Note that the DtN map $\Lambda_{\sigma,D}$ is usually represented with respect to an orthonormal basis, most commonly spherical harmonics or a trigonometric cosine/sine basis for real-valued injections. Nevertheless, in practice pairwise injections are collected for most systems, which is the case for the KIT4 system \citep{kourunen2008suitability} described in Section~\ref{sec:experimental_setup}. Thus, first one has to transform pairwise injections and measurements, by change of basis, into the trigonometric basis to represent the NtD-map $\Lambda_{\sigma,N}$. The corresponding current patterns (and basis) $I^{(k)}\in\R^L$  for $L$ and $k=1,\dots,L-1$ are given by
\begin{equation}
    I^{(k)}_l= 
    \left \{ \begin{array}{lr}
    \displaystyle \cos\left((k+1)\theta_l/2\right),   &\mbox{for odd }k,  \hspace{0.1 cm} \, \\
    \displaystyle \sin\left(k\theta_l/2\right),  &\mbox{ for even } k.
    \end{array}\right.
\end{equation}
The estimated NtD map $\Lambda_{\sigma,N}$ is then used to approximate the continuum data \citep{hyvonen2009approximating,hauptmann2017approximation}, and by inversion we obtain the DtN map $\Lambda_{\sigma,D}$ needed for the D-bar method. 
Below we  outline the classic D-bar method for $\sigma\in C^2(\Omega)$, with $\sigma \geq c>0$ in $\Omega$, and $\sigma\equiv 1$ in the neighbourhood of the boundary $\Gamma$. Consider the embedding of $\mathbb{R}^2$ in the complex plane, and identifying a planar point $x=(x_1,x_2)$  with the  complex number $x_1+{\rm i}x_2$, and the product $kx$ denotes complex multiplication. See the survey by \cite{MuellerSiltanen:2020} for further details.

First, the conductivity equation \eqref{eqn:eit} is transformed into a Schr\"odinger-type equation by substituting $\widetilde{u}=\sqrt{\sigma}u$, setting $q=\Delta \sqrt{\sigma}/\sqrt{\sigma}$, and extending $\sigma\equiv 1$ outside $\Omega$. Then we obtain
\begin{equation}\label{eqn:Schrodinger}
    (-\Delta + q(x))\widetilde{u}(x) = 0,\quad  \mbox{in }\mathbb{R}^2.
\end{equation}
Next, we define a class of special solutions of \eqref{eqn:Schrodinger} due to \cite{Faddeev:1966}, i.e., complex geometrical optics (CGO) solutions $\varphi(x,k)$, which depend on a complex parameter $k\in\C \setminus \{0\}$ and $x\in\mathbb{R}^2$. These exponentially behaving functions are key to the reconstruction. Specifically, given $q\in L^p(\mathbb{R}^2), \, 1<p<2$, the CGO solutions $\varphi(x,k)$ are defined to be the solutions to
\[
(-\Delta + q(x))\varphi(\cdot,k) = 0,\quad \mbox{in }\mathbb{R}^2,
\]
satisfying the asymptotic condition $e^{-{\rm i}kx}\varphi(x,k)-1 \in W^{1,\tilde{p}}(\R^2)$ with $2<\tilde{p}<\infty$. These solutions are unique for $k\in\C\setminus\{0\}$ \cite[Theorem 1.1]{Nachman:1996}. The  D-bar method recovers the conductivity $\sigma$ from the knowledge of the CGO solutions $\mu(x,k)=e^{-{\rm i}kx}\varphi(x,k)$ at the limit $k\to 0$ \cite[Section 3]{Nachman:1996} 
\begin{equation*}
\lim_{k\to 0}\mu(x,k)    =\sqrt{\sigma},\quad x\in\Omega.
\end{equation*}
Numerically, one can substitute the limit by $k=0$ and evaluate $\mu(x,0)$.
The reconstruction of $\sigma$ uses an intermediate object called the (non-physical) scattering transform $\mathbf{t}$, defined by 
\begin{equation*}\label{eqn:scatTrafo}
    \mathbf{t}(k) = \int_{\mathbb{R}^2}e_k(x)\mu(x,k)q(x){\rm d}x,
\end{equation*}
with $e_k(x):=\exp({\rm i}(kx+\bar k\bar x))$, where over-bar denotes complex conjugate. Since $\mu$ is asymptotically close to one, $\mathbf{t}(k)$ is similar to the Fourier transform of $q(x)$.  We obtain $\mu$ by solving the name-giving D-bar equation
\begin{equation}\label{eqn:Dbar}
\bar\partial_ k \mu(x,k)=\frac{1}{4\pi\bar{k}}\mathbf{t}(k)e_{-k}(x)\overline{\mu(x,k)},\quad k\neq 0,
\end{equation}
where $\bar\partial_k =\frac12 (\frac{\partial}{\partial k_1}+{\rm i}\frac{\partial}{\partial k_2})$ is known as the D-bar operator.
To solve \eqref{eqn:Dbar}, the scattering transform $\mathbf{t}(k)$ is required, which we cannot measure directly in the experiment, but $\mathbf{t}(k)$ can be represented using the DtN map $\Lambda_{\sigma,D}$. Indeed, using Alessandrini's identity \citep{Alessandrini}, we get 
\begin{equation} \label{eq:scatTrafo_comp}
\mathbf{t}(k)=\int_{\Gamma} e^{{\rm i}\bar{k}\bar{x}}(\Lambda_{\sigma,D}-\Lambda_{1,D}) \varphi(x,k) {\rm d}s.
\end{equation}
Note that $\Lambda_{1,D}$ can be analytically computed, and only $\Lambda_{\sigma,D}$ needs to be obtained from the measurement.  One can either compute the scattering transform $\mathbf{t}(k)$ using a Born approximation $\varphi\approx e^{{\rm i}kx}$, leading to a linearized approximation, or solve a Fredholm boundary integral equation of the second kind using Faddeev's Green's function as well as the measured difference operator $\Lambda_{\sigma,D}-\Lambda_{1,D}$ \citep{Nachman:1996}. In the numerical experiments, we employ the full nonlinear algorithm.

The resulting D-bar algorithm is highly parallelizable in each step. First, the computation of the CGO solutions $\varphi(x,k)$ and $\mathbf{t}(k)$ are both independent of $k$. Second, the solutions of \eqref{eqn:Dbar} are also independent for each $x\in\Omega$ and one can efficiently parallelize over $x$. This leads to real-time implementations~\citep{dodd2014real}, which is especially relevant for time-critical applications, e.g., monitoring purposes.  

Note that the algorithm assumes infinite precision and exact data. When the measured data is noisy and comes from a finite number of electrode measurements, the resulting DtN map $\Lambda_{\sigma,D}$ is inaccurate, and the computation of $\mathbf{t}(k)$ becomes exponentially unstable for $|k|>R$ for some radius $R$. Thus, in practice, we restrict the computation to a certain frequency range in order to stably compute $\mathbf{t}(k)$. We choose $R=4$ throughout this study. The overall procedure is summarized in Algorithm \ref{alg:Dbar}.

\begin{algorithm}\small
\caption{D-bar algorithm using $\mathbf{t}^{\exp}$}
\label{alg:Dbar}
\KwIn{ $\Lambda_{\sigma,D}$ and $R$}
\KwResult{Regularised reconstruction of $\sigma$}
Compute analytic $\Lambda_{1,D}$\;
Obtain $\varphi(x,k)$ for $|k|<R$ from boundary integral equation; \\
Evaluate $\mathbf{t}(k)$ for $|k|<R$ by \eqref{eq:scatTrafo_comp}\;
Solve the D-bar equation \eqref{eqn:Dbar}\;
Obtain $\sigma(x) = \mu(x,0)^2$ for $x\in\Omega$\;
\end{algorithm}

\subsubsection{Sampling methods}
 \label{SM}

Sampling methods are a class of qualitative methods in which a set of test points is selected to determine whether each point belongs to the inclusion support. Specifically, the unknown $\sigma \colon \Omega \to \mathbb{R}$ consists of a known constant background $\sigma_0$ and unknown inclusions, modeled as a subset $D \subset \Omega$. In a sampling method, one constructs an indicator function to indicate for each test point $x \in \Omega$, whether it lies within $D$. These methods are typically tailored to the structure and properties of the specific inverse problem, and include  linear sampling method \citep{kirsch1998, c2001, Somer2001}, Multiple Signal Classification \citep{c2001, Lech2015}, and direct sampling method \citep{ItoJinZou:2012,chow2014direct,ItoJinWang:2025}. All of them were initially developed for inverse scattering problems and then adapted to EIT.

\paragraph*{Factorization method}
The factorization method due to \cite{HB2003} is one classical sampling technique for the continuum model \eqref{eqn:eit} of EIT. In the method, one factorizes the difference between the NtD operators of the true conductivity $\sigma$ and the background $\sigma_0$, and constructs a function $h_z$ for each point $z \in \Omega$. Then $z$ lies within the inclusion $D$ if and only if $h_z$ belongs to the range of the operator $(\Lambda_{\sigma,N} - \Lambda_{\sigma_0,N})^{1/2}$.
By the Picard criterion, we deduce
\[
z \in D \quad \Longleftrightarrow \quad \sum_{n=0}^\infty \frac{\langle h_z,v_n\rangle^2}{\lambda_n} < \infty,
\]
where $(\lambda_n)$ and $(v_n)$ are the eigenvalues and eigenfunctions of the difference operator $\Lambda_{\sigma,N} - \Lambda_{\sigma_0,N}$, respectively. 

\paragraph*{Direct sampling method}
The direct sampling method (DSM) \citep{chow2014direct} is also well-known for identifying conductivity anomalies using the continuum model \eqref{eqn:eit}. 
Using only one single pair of Cauchy data on the boundary $\Gamma$, the DSM constructs a family of probing functions $\{\eta_{x,d_x}\}_{x \in \Omega, d_x \in \mathbb{R}^d} \subset H^{2\gamma}(\Gamma)$, for some $\gamma \geq 0$, and defines an index function
\begin{equation*}
	\mathcal{I}(x,d_x):=\frac{\langle \eta_{x,d_x},u_\sigma-u_{\sigma_0} \rangle_{\gamma,\Gamma}}{\Vert u_\sigma-u_{\sigma_0}\Vert_{L^2(\Gamma)}|\eta_{x,d_x}|_Y},\quad x\in \Omega,
\end{equation*}
where $u_\sigma$ and $u_{\sigma_0}$ are the solutions of problem \eqref{eqn:eit} for $\sigma$ and $\sigma_0$, respectively, both with the current $j$. The notation $|\cdot|_Y$ denotes the $H^{2\gamma}(\Gamma)$ seminorm and the duality product $\langle f,g \rangle_{\gamma,\Gamma}$ is defined by 
\begin{equation*}
\langle f,g\rangle_{\gamma,\Gamma}=\int_{\Gamma}(-\Delta_{\Gamma})^{\gamma}fg{\rm d}s=\langle (-\Delta_{\Gamma})^{\gamma}f, g\rangle_{L^2(\Gamma)},	
\end{equation*}
where $-\Delta_{\Gamma}$ denotes the Laplace-Beltrami operator, and $(-\Delta_{\Gamma})^\gamma$ its fractional power defined via spectral calculus. More precisely, let the Cauchy difference function $\varphi$ be defined by the solution of
\begin{equation}\label{Cauchy_data}
\left\{\begin{aligned}	
-\Delta \varphi&=0, \quad&&\text{in }\Omega,\\
\frac{\partial\varphi}{\partial n}&=(-\Delta_{\Gamma})^\gamma(u_\sigma-u_{\sigma_0}), \quad &&\text{on }\Gamma,\\ 
\int_{\Gamma} \varphi {\rm d}s&=0,
\end{aligned}\right.
\end{equation}
i.e., the EIT forward problem for the background conductivity with the applied current $j$ replaced by  $u_\sigma-u_{\sigma_0}$. Then the index function $\mathcal{I}(x,d_x)$ is given by
\begin{equation*}
	\mathcal{I}(x,d_x):=\frac{d_x\cdot\nabla \varphi(x)}{\Vert
    \phi_\sigma-\phi_{\sigma_0}\Vert_{L^2(\Gamma)}|\eta_{x,d_x}|_Y},\quad x\in \Omega.
\end{equation*}
The unit vector $d_x = \nabla \varphi(x)/\|\nabla \varphi(x)\|_{L^2(\Gamma)}$ was chosen to carry out the numerical experiments by \cite{chow2014direct} to prevent the direction $d_x$ from being orthogonal to $\nabla \varphi(x)$ and accidentally removing existing inclusions.

Ideally the index function $\mathcal{I}(x,d_x)$ should assume large values at points near / inside the inclusions and relatively small values for points far away from $D$.  This relies heavily on the choice of the Sobolev dual product $\langle\cdot,\cdot\rangle_{\gamma,\Gamma}$. Very recently, this approach was extended by \cite{ItoJinWang:2025}, by incorporating an iterative mechanism, so as to achieve better resolution and numerical stability. However, these developments are for the continuum model \eqref{eqn:eit}. Below we propose a deep learning version of the DSM for the CEM. 

 \section{Learned reconstruction methods} 
\label{sec:DL}
The development of deep learning techniques has significantly advanced EIT reconstruction in recent years. Deep neural networks (DNNs) address the ill-posed nature of EIT by choosing specific network architectures and by learning nonlinear mappings between input data and reconstruction outputs. Recent studies demonstrate notable improvements in key metrics, e.g., noise robustness, structural fidelity, and resolution enhancement, thereby significantly broadening EIT's applicability in biomedical diagnostics and nondestructive testing. The aim of learned methods is to replace parts or even the entire reconstruction process by DNNs trained on available (paired) training data. These methods differ in the parts that are replaced by DNNs, and how the DNNs are trained. 

All deep learning approaches discussed in this study utilize convolutional neural networks (CNNs), which are particularly well-suited for pixel-based image data. However, in EIT, the forward operator is discretized using FEM meshes, which are typically composed of triangular elements. This mesh-based representation is incompatible with CNNs, necessitating an interpolation step to transform the mesh data into a pixel-grid format. \cite{Herzberg:2021,herzberg2023domain} and \cite{toivanen2025graph} address this limitation using graph convolutional networks to process mesh-based data directly, eliminating the need for interpolation. Additionally, domain-adapted network architectures, e.g., neural operators \citep{azizzadenesheli2024neural}, offer a promising research avenue by learning mappings between function spaces, enabling efficient handling of data on irregular or mesh-based domains.

In the following we survey different supervised and unsupervised learned reconstruction methods, and the specific approaches  used in the comparative study. Roughly these approaches can be grouped into 
\begin{itemize}
\item Fully learned approaches
\item Post-processing methods
\item Deep direct sampling method
\item Learned iterative methods
\item Unsupervised methods
\end{itemize}
Recently, also deep generative models have been applied to EIT; see the review by \cite{WangXuZhou2024}. For example, \cite{alberti2024manifold} learn a mixture of variational autoencoders \citep{KingmaVae2014} to approximate the data manifold. Also, \cite{seo_harrach_2019} propose to use variational autoencoders and to exploit the low dimensional latent space for lung EIT.
\cite{Denker2024EIT} employ score-based models \citep{SongScore2011} to model the posterior distribution.
However, these methods are not included in the study.


\subsection{Fully-learned approaches} 
\label{sec:fully_learned}
In the existing literature, a substantial body of research focuses on fully-learned approaches by designing DNN architectures $\mathcal{R}_\theta$ that directly approximate the inverse mapping from boundary voltage measurements $\mathbf{U}^\delta$ to the conductivity distributions $\sigma$, such that $\sigma\approx \mathcal{R}_\theta(\mathbf{U}^\delta)$, based on supervised learning over paired training datasets. Early approaches by \cite{adler1994neural} employ a single linear layer to learn the relationship between the mapping and conductivity. \cite{li2017image} proposed a four-layer architecture built upon stacked autoencoders and a logistic regression layer. \cite{tan2018image} followed with a convolutional model inspired by LeNet, incorporating pooling and dropout layers to enhance the robustness. More recent developments include hybrid designs such as the FC-UNet proposed by \cite{9128764}, which first uses a fully connected layer to convert voltage data into image features before applying a UNet for final reconstruction. \cite{yang2023eit} introduced a DenseNet model featuring multiscale convolution to further capture fine conductivity structures. Compact and low-rank-aware architectures have  been proposed by \cite{fan2020solving}, targeting both forward and inverse problems in 2D and 3D. \cite{huang2019improved} combine radial basis function models for initial estimates with UNet-based refinement. Meanwhile, variational autoencoder-based approaches, see e.g. \cite{seo_harrach_2019}, explore latent space representations of the conductivity to facilitate dimensionality reduction. 

One prototypical example of a fully-learned approach is the FC-UNet \citep{9128764}. We implement a variation of the method in the experimental evaluation. The FC-UNet directly learns a mapping from the measurements to reconstructions with a linear fully connected layer and a UNet \citep{ronneberger2015u}, eliminating the need for any physical modeling. 
We parametrize the reconstruction operator $\mathcal{R}_\theta$ as
\begin{align*}
    \mathcal{R}_\theta(\mathbf{U}^\delta) := \mathcal{G}_\theta(\mathcal{S}(\mathbf{W} \mathbf{U}^\delta)),
\end{align*}
where $\mathbf{W}$ is a learnable matrix, mapping the measurements to an initial image, and $\mathcal{G}_\theta$ is implemented as a UNet. The parameter count of the initial matrix $\mathbf{W}$ grows quadratic with the image resolution. For example, with a resolution $N=128$ and the KIT4 setting ($K=76$ and $L=16$) the matrix $\mathbf{W} \in \R^{128^2 \times KL}$ already has about $20$M parameters. The high parameter count poses major computational and memory challenges during training. To partially address these challenges, we use a smaller initial image resolution ($N=64$), followed by a bilinear upsampling $\mathcal{S}:\R^{64^2} \to \R^{256^2}$ to match the target resolution of the UNet.
In the original implementation of the FC-UNet \citep{9128764} the authors use the ReLU  after the initial linear layer to constrain the input to the UNet to be nonnegative. In our experiments, we find that this additional operation is not necessary and we remove it for simplicity. 

We employ a two-step approach to train the FC-UNet. First, by initializing the matrix $\mathbf{W}$  to zero, we train it via the mean squared error loss 
\begin{align}
    \min_\mathbf{W} \sum_{i=1}^n \| \mathcal{S}(\mathbf{W} \mathbf{U}^{\delta,(i)}) - \sigma^{(i)} \|_2^2.
\end{align} 
This training procedure estimates an optimal linear reconstruction operator $\mathbf{W}$ with respect to the least-squares loss. After pre-training the matrix $\mathbf{W}$, we train the entire network (including $\mathbf{W}$) by minimizing 
\begin{align}
        \min_{\mathbf{W}, \theta} \sum_{i=1}^n \| \mathcal{G}_\theta(\mathcal{S}(\mathbf{W}  \mathbf{U}^{\delta,(i)})) - \sigma^{(i)} \|_2^2.
\end{align}
We use a total of $110$ epochs, with $10$ epochs for the pre-training of $\mathbf{W}$ and $100$ epochs for the full network. The UNet has an input resolution of $256$ and employs $4$ downsampling steps. We use attention layers at resolution $32$ and $16$ following \cite{dhariwal2021diffusion}. The total network has \num{11486625} parameters, where \num{4980736} are used by the initial linear layer. We employ the Adam optimizer \citep{KingmaBa:2015} with a cosine annealing learning rate, starting at $1\mathrm{e}{-4}$ and reducing to $1\mathrm{e}{-7}$ with a batch size of $32$. For the final evaluation we pick the model with the lowest validation loss. 

\subsection{Post-processing methods} 
\label{sec:lin_post}
To leverage the stability and interpretability of analytic methods, many recent studies have explored hybrid strategies based on post-processing where trained neural networks are employed to refine or enhance results from model-based reconstructions. Formally, these methods learn $\sigma\approx \mathcal{G}_\theta(\mathcal{R}(\mathbf{U}^\delta))$, where $\mathcal{R}$  is a classical (non-learned) reconstruction operator. One prominent example is the Deep D-bar method~\citep{hamilton2018deep}, which employs the D-bar method for the initial image reconstruction followed by a UNet for enhancement. Extensions of this strategy include deep versions of Calder\'{o}n’s method \citep{cen2023electrical,SunZhongWang:2023,LiShinZhou:2025}, the domain-current method \citep{wei2019dominant}, iterative solvers like Gauss-Newton \citep{martin2017post} and conjugate gradient \citep{zhang2022v} and multi-scale frequency enhanced deep D-bar method \cite{CaoDingZhang:2025}. The Deep DSM (DDSM) due to \cite{guo2021construct} is a further innovation different from post-processing, leveraging Cauchy difference functions (instead of approximations to the inclusions) derived from the direct sampling method as inputs to a CNN. 

We implement two post-processing approaches: the deep D-bar method and linearized post-processing. These methods share the neural network architecture for refinement but differ in the choice of the initial reconstruction. We use the same UNet architecture as for the FC-UNet, including attention layers and a total parameter count of $\num{6305185}$

In the deep D-bar method \citep{hamilton2018deep}, we first obtain a stable regularized reconstruction using the D-bar method, denoted by $\mathcal{R}_{\overline{\partial}}$ (see Algorithm \ref{alg:Dbar}). This step maps the measured DtN-map $\Lambda_{1,D}^{(i)}$ for sample $i$ to an approximate reconstruction, i.e., $\sigma^{(i)}\approx\mathcal{R}_{\overline{\partial}}(\Lambda_{1,D}^{(i)})$. The UNet reconstructs the pixel values on a rectangular grid. One distinct strength of the D-bar method in the context is that the reconstructions can be computed for any $x\in\Omega$ (even for any $x\in\R^2$). \cite{hamilton2019beltrami}  extended this capability to train domain-shape-independent networks. Initially, the D-bar reconstructions are generated on the $[-1,1]^2$ pixel grid with $128^2$ pixels. Even though some points on this grid lie outside the measurement domain, the reconstructions at these points can contain valuable information, enhancing the post-processing stage.

In the linearized post-processing method, the input to the UNet $\mathcal{G}_\theta$ is the linearized reconstruction $\DeltaSigmaHat$, cf.~\eqref{eq:lin_rec_one_step}.  Since the linearized reconstruction is defined on a mesh while the UNet operates on a pixel grid, a linear interpolation step is incorporated to map the reconstruction onto the pixel grid.

Given the initial reconstruction $\sigma_\text{reco}$, either given by $\mathcal{R}_{\overline{\partial}}(\Lambda_{1,D})$ or $\DeltaSigmaHat$, the CNN is trained to minimize the mean squared error over the paired training data $\{(\sigma_{\rm reco}^{(i)},\sigma^{(i)})\}$:
\begin{align}
        \min_{\theta} \sum_{i=1}^n \| \mathcal{G}_\theta(\sigma_\text{reco}^{(i)})) - \sigma^{(i)} \|_2^2,
\end{align}
to match the ground truth conductivity distribution. To minimize the loss, we employ the Adam optimizer \citep{KingmaBa:2015}. We train the model for 200 epochs with a cosine learning rate decay with an initial learning rate of $1\mathrm{e}{-4}$, decreasing to a final learning rate of $1\mathrm{e}{-5}$. The model selected for final evaluation is the one achieving the lowest mean squared error on the validation set. Note that once the DNN $\mathcal{G}_\theta$ is trained, for each test data $\Lambda_{\sigma,D}^{*}$, deploying the method requires only feeding the initial reconstruction through the neural network, which is computationally very efficient.

\subsection{Deep direct sampling method}
\label{sec:deepdsm}


Inspired by the direct sampling method (DSM) \citep{chow2014direct}, \cite{guo2021construct} proposed the Deep Direct Sampling Method (DDSM), which employs multiple pairs of Cauchy data to train a CNN-based UNet network \citep{ronneberger2015unet} to learn a relationship between the Cauchy difference functions $\varphi_i$ and the true inclusion distribution. 
More specifically, the DDSM constructs and trains a UNet $\mathcal{G}_\theta$ such that 
\begin{equation}\label{CNN}
	\sigma\approx\mathcal{G}_\theta(\varphi_1,\varphi_2,...,\varphi_N),
\end{equation}   
where $\{\varphi_i\}_{i=1}^{N}$ correspond to $N$ pairs of Cauchy data  $\{j_\ell,\Lambda_{\sigma,N} j_\ell\}_{\ell=1}^N$, see \eqref{Cauchy_data}. The following theorem \citep[Theorem 4.1]{guo2021construct} provides the mathematical foundation for the DDSM.
\begin{thm}
Let $\{j_\ell\}_{\ell=1}^\infty$ be a fixed orthonormal basis of $H^{-1/2}(\Gamma)$. Given
an arbitrary $\sigma$ such that $\sigma>\sigma_0$ or $\sigma<\sigma_0$, let $\{j_\ell,\Lambda_{\sigma,N} j_\ell\}_{\ell=1}^\infty$ be the Cauchy data pairs and let $\{\varphi_\ell\}_{\ell=1}^\infty$ be the corresponding Cauchy difference functions. Then the inclusion distribution $\sigma$ can be uniquely determined from $\{\varphi_\ell\}_{\ell=1}^\infty.$
\end{thm}
The idea of the DDSM has been further extended to a transformer-based approach by \cite{GuoCaoChen:2023}, using the  harmonic extension as different input channels, which is inspired by the DSM. The authors employed learnable non-local kernels, in which direct sampling is recast to a modified attention mechanism. 

To the best of our knowledge, the (D)DSM for the CEM has not been studied so far. Motivated by the method developed by \cite{guo2021construct}, we propose a version of the DDSM for the CEM by training a UNet network, as in \eqref{CNN}, to learn the relationship between the Cauchy difference functions $\{\varphi_i\}_{i=1}^{L}$ and the sought-after conductivity $\sigma$. More precisely, in the CEM version of the DDSM, each function $\varphi_i$ corresponds to the first entry of the pair $(\varphi_i,V)\in H^1(\Omega)\oplus\mathbb{R}_\diamond^L$, which is the solution of the following PDE
\begin{equation*}
\left\{\begin{aligned}
-\Delta \varphi_i&=0,\quad &&\mbox{in }\Omega,\\
\varphi_i+z_\ell\frac{\partial \varphi_i}{\partial n}&=V_l,\quad &&\mbox{on } e_\ell, \ell=1,2,\ldots,L,\\
\int_{e_\ell}\frac{\partial \varphi_i}{\partial n}{\rm d}s& =(U^{\delta,(i)}_\sigma-U^{(i)}_{\sigma_0})_\ell,\quad &&\ell=1,2,\ldots, L,\\
\frac{\partial \varphi_i}{\partial n}&=0,\quad &&\mbox{on } \Gamma\backslash\cup_{\ell=1}^Le_\ell,
\end{aligned}\right.
\end{equation*}
with $V = (V_1,\dots,V_L)$. In addition, $U^{\delta,(i)}_\sigma,U^{(i)}_{\sigma_0} \in \mathbb{R}^L$ are the (measured) potentials at the electrodes for the conductivity $\sigma$ and the background $\sigma_0$, respectively. Note that similarly to the method of Guo and Jiang, the function $\varphi_i$ corresponds to (part of) the solution of the PDE that models the forward problem of EIT, with the background constant conductivity, and with the current pattern replaced by the difference $U^{\delta,(i)}_\sigma-U^{(i)}_{\sigma_0}$, see \eqref{Cauchy_data} and \eqref{eqn:cem}. Formally it lifts the electrode voltage difference to the domain $\Omega$.

The final model is based on the UNet architecture with an input and output resolution of $64\times 64$px. The model has $16$ input channels, one for each (discretised) difference function $\{\varphi_i\}_{i=1}^{L}$. It has \num{24006273} trainable parameters. 

\subsection{Learned iterative methods} 
\label{sec:unrolled}
One prominent class of hybrid methods relies on unrolling iterative solvers into DNNs. These architectures learn parameterized update rules by mimicking optimization procedures such as ADMM or Gauss-Newton iterations. Originally, learned iterative methods were proposed for image reconstruction in linear inverse problems, with notable applications in MRI and CT \citep{adler2017solving,adler2018learned,hammernik2018learning}. Extensions to EIT include the MMV-Net \citep{ZhouYang:2022}, which unfolds ADMM steps for multi-frequency EIT, and unrolled Gauss-Newton method (\cite{Herzberg:2021} and \cite{colibazzi2023deep} for learning Gauss-Newton updates and proximal operators, respectively).
Quasi-Newton updates via learned SVD approximations have also been proposed \citep{smyl2021efficient}.  \cite{AlbertiRatti:2025} unroll  the iterative proximal regularized Gauss-Newton method with graph neural networks (GNNs) for multi-frequency EIT. These approaches offer a balance between physical modelling and data-driven learning. 

In this study we investigate the learned Gauss-Newton method: we unroll a (fixed) number of Gauss-Newton updates
\begin{align}
    \sigma_{k+1} := \sigma_k + \delta \sigma_k, \quad k=1, \dots, K,
\end{align}
where $\delta \sigma_k$ is given by the Gauss-Newton update step. Then, the additive update $\delta\sigma$ is replaced with a learned update using a neural network
\begin{align}
    \sigma_{k+1} = \mathcal{G}_{\theta_k}(\sigma_k,\delta \sigma_k), \quad k=1, \dots, K,
\end{align}
where each subnetwork has individual parameters $\theta_k$. Similar to \cite{Herzberg:2021} we employ Gauss-Newton updates with a smoothed TV regularization, cf. \eqref{eq:tv_update} in Section~\ref{sec:variational_methods}. Since the Gauss-Newton update step is given on the FEM mesh, \cite{Herzberg:2021} employ GNNs for all subnetworks $\mathcal{G}_{\theta_k}$ to directly process the mesh representation. However, instead of employing GNNs, we utilize CNNs as in \cite{mozumder2021model}, which requires additional interpolation steps both before and after applying the FEM model. The interpolation layers map the mesh to a regular grid suitable for CNNs and then map the output of the network back to the FEM domain. The full model, referred to as Deep GN-TV, is trained iteratively using a greedy training strategy \citep{hauptmann2018model}: each subnetwork is trained iteratively, i.e., for $k=1, \dots, K$, we minimize the loss
\begin{align}
    \min_{\theta_k} \sum_{i=1}^n \| \sigma^{(i)} - \mathcal{G}_{\theta_k}(\sigma_{k-1}^{(i)}, \delta \sigma_{k-1}^{(i)}) \|_2^2,
\end{align}
where $\sigma_{k-1}^{(i)}$ is the prediction from the previous subnetwork for the $i$-th example and $\delta \sigma_{k-1}^{(i)}$ is the GN-TV update. After training the $k$-th subnetwork, all weights in the subnetwork are frozen and the next subnetwork $k+1$ is trained. Greedy training results in iteration-wise optimality, which serves as an upper bound to end-to-end optimality. This training method separates the training of the subnetworks and the application of the FEM model, which reduces the computational complexity. Further, we do not have to back-propagate through the FEM model and through the construction of the Jacobian. 
In our implementation, we unroll GN-TV for $6$ iterations and each subnetwork $\mathcal{G}_{\theta_k}$ is implemented as a small UNet with $3$ downsampling operations. Each subnetwork has $\num{124265}$ trainable parameters. The model has $\num{745590}$ parameters.

Learned iterative methods often achieve the best performance in practice~\citep{arridge_acta}. However, they have a significant limitation in terms of the computational cost. For each example and iteration, they require the GN-TV update step, which involves computing the Jacobian and thus substantially increases the computational complexity. In contrast, post-processing and fully-learned methods are relatively inexpensive to deploy. For example, the linearized post-processing approach requires only one single Jacobian of the constant background (which can be reused for multiple inputs), while the FC-UNet requires only a single forward pass through the DNN.

\subsection{Unsupervised methods}\label{UM}

Despite their effectiveness, supervised models  depend heavily on the quality, size and representativeness of the training data. Due to the challenge associated with acquiring large-scale paired training data under varied settings, most current datasets in EIT are synthetically generated. While training on simulated phantoms enables flexibility, generalization can be limited: These methods tend to perform well only when the test data closely matches the training distribution. In contrast, robustness against domain shift, geometry variations, and realistic noise remains a pressing challenge in deep learning-based image reconstruction techniques \citep{Antun:2020}. In this survey we focus on supervised learned methods, i.e., on deep learning concepts that explicitly exploit the data structure. However, there do exist various unsupervised concepts, that have been applied to the EIT problem. We will briefly review the state of research but do not include such methods in the subsequent numerical comparison.

To reduce the need of ground-truth data, several physics-driven unsupervised and weakly supervised strategies have been developed, incorporating physical constraints directly into the learning process. Physics-Informed Neural Networks~\citep{raissi2019physics} represent a major milestone in this direction. \cite{bar2021strong} proposed using DNNs to simultaneously represent the conductivity $\sigma$ and multiple voltage potentials $\{u_j\}$, training the model to satisfy the governing PDE and boundary conditions in a strong form (in the continuum model). Similar approaches combine data-driven learning with energy-based models to enhance robustness and convergence \citep{pokkunuru2022improved}. 
One unsupervised direction is the Deep Image Prior \citep{ulyanov2018deep}, applied to EIT by \cite{liu2023deepeit}, in which a CNN is trained per instance by optimizing image quality against measurement consistency, using no external data. \cite{ChenLiu:2025} developed an approach for high-contrast EIT by combining latent surface representation and shape optimization. While being computationally intensive, these physics-driven models and their variants offer high adaptivity and robustness under distributional shift, since they do not rely on paired datasets. 

\section{Experimental setup}   
\label{sec:experimental_setup}
Due to the lack of large paired datasets for training learned reconstruction methods in EIT, we rely on simulated data. We create a dataset of simulated elliptical inclusions. This dataset is used to train the learned reconstructors and to choose the hyperparameters (e.g., regularization parameters and number of iterations) for the classical reconstruction methods. For testing, we utilize a held-out subset of these simulated ellipses and further evaluate the methods on phantoms featuring out-of-distribution inclusions, e.g., rectangles and triangles. Finally, all methods are evaluated on real measured data from the KIT4 dataset~\citep{hauptmann2017open}. The KIT4 dataset does not provide ground-truth conductivity distribution, only photos of the watertank containing different inclusions. We make use of these photographs and create hand-annotated segmentation masks, which are then used in the evaluation routine. In this section, we first define the evaluation metrics used, and then describe the datasets and simulation setup.

\subsection{Evaluation scores}
\label{sec:evaluation_scores}
We evaluate the reconstruction methods using a range of scores that highlight different aspects of reconstruction quality. All evaluation scores are calculated on the FEM mesh. For CNN-based network architectures, this requires an interpolation step, mapping the the pixel-based image to the FEM mesh.

\noindent \textbf{Relative $L^1$/$L^2$-error } First, we use two global metrics, the \textit{relative $L^1$-error} and the \textit{relative $L^2$-error}, to assess the overall reconstruction accuracy. Given the ground truth conductivity distribution $\sigma_\text{gt}$ and the reconstruction $\sigma_\text{rec}$ and the relative $L^1$-error and the relative $L^2$-error are defined respectively by
\begin{align*}
    \text{L1rel}(\sigma_\text{gt}, \sigma_\text{rec}) &= \frac{\int_\Omega | \sigma_\text{gt} - \sigma_\text{rec} | {\rm d}x}{\int_\Omega | \sigma_\text{gt} | {\rm d}x},\\
    \text{L2rel}(\sigma_\text{gt}, \sigma_\text{rec}) &= \frac{\int_\Omega ( \sigma_\text{gt} - \sigma_\text{rec} )^2 {\rm d}x}{\int_\Omega \sigma_\text{gt}^2 {\rm d}x}.
\end{align*}
\noindent \textbf{Dice Score } Next, we compute a task-based score, i.e., dice score, to evaluate the accuracy of the localization of inclusion, similar to~\cite{Mikko}. The motivation behind dice score is that while some methods may overestimate or underestimate the intensity, they may still perform well in localizing the inclusions. Using Otsu's algorithm~\citep{otsu1975threshold}, the conductivity distribution is segmented into resistive and conductive inclusions and background regions. We then compute the \textit{dice score} (DS) as the mean dice coefficient over the three classes, where the dice coefficient is defined as 
\begin{align}
    \text{DSC} = \frac{2 \text{TP}}{2 \text{TP} + \text{FP} + \text{FN}},
\end{align}
with true positive (TP), false positive (FN) and false negative (FN). The dice coefficient is between 0 and 1, where 1 corresponds to a perfect segmentation. Note that the Dice score may not reflect image quality accurately, due to the involved segmentation.

\noindent \textbf{Dynamic Range } To measure whether the intensity values are overestimated or underestimated by a reconstruction method, we compute the \textit{dynamic range} (DR) \citep{Herzberg:2021} as 
\begin{align}
\text{DR}(\sigma_\text{gt}, \sigma_\text{rec}) = \frac{\max(\sigma_\text{rec}) - \min(\sigma_\text{rec})}{\max(\sigma_\text{gt}) - \min(\sigma_\text{gt})}.     
\end{align}
A dynamic range of $1$ means that the intensity range of the ground truth was reconstructed perfectly.

\noindent \textbf{Measurement Error } Finally, we compute the \textit{measurement error}. It measures the error between the predicted potential at the electrodes $\mathbf{F}(\sigma_\text{rec}) \mathbf{I}$ and the measurements $\mathbf{U}^\delta$. We compute it as 
\begin{align}
    \text{MeasErr}(\sigma_\text{rec}, \mathbf{U}^\delta) = \frac{\| \mathbf{F}(\sigma_\text{rec}) \mathbf{I} - \mathbf{U}^\delta \|_2^2}{\|\mathbf{U}^\delta \|_2^2}.
\end{align}
The measurement error coincides with the data fitting of the reconstruction $\sigma_\mathrm{rec}$.

\subsection{Simulation setting}
\label{sec:sim_data}
\begin{figure}
    \centering
    \includegraphics[width=1.0\linewidth]{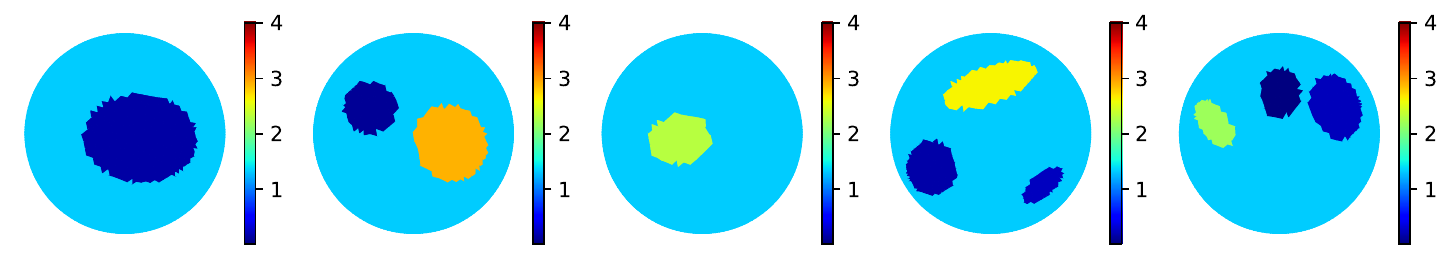}
    \rule{0.8\linewidth}{0.4pt}
    \includegraphics[width=1.0\linewidth]{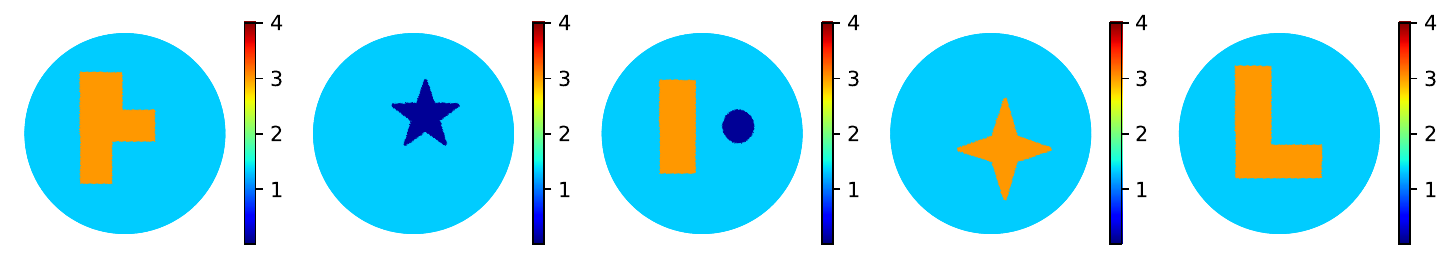}
    \caption{Exemplary ground truth phantoms of the simulated dataset. First row: In-distribution samples with elliptical inclusions. Second row: Out-of-distribution phantoms.}
    \label{fig:ood_phantoms}
\end{figure}

We create a synthetic dataset of conductivity distributions, following the simulation protocol described in~\cite{Tanyu2025}. Each phantom contains $1$ to $3$ non-overlapping elliptical inclusions with a uniform background with a conductivity value $1.31$, matching the KIT4 system. The conductivity values of the inclusions are sampled independently, with lower values sampled from $\mathcal{U}[0.01, 0.3]$ and higher values sampled from $\mathcal{U}[2.0, 3.0]$. The dataset is divided into training (2000 pairs), validation (200 pairs), and test (100 pairs) subsets. In addition to these elliptical inclusions, we create out-of-distribution phantoms with different shapes. See Figure~\ref{fig:ood_phantoms} for both in-distribution and out-of-distribution phantoms.

Measurements were simulated according to the KIT4 setup \citep{hauptmann2017open}, i.e., a circular domain with $16$ equidistant electrodes covering $45\%$ of the boundary. 
We utilize two meshes with triangular elements: a coarse mesh with $5248$ elements, and a dense mesh with $8072$ elements. For both meshes, we align the mesh points on the boundary with the extend of the electrode, such that each electrode is perfectly covered by a fixed number of mesh elements. The dense mesh was used in the FEM solver, see Section \ref{sec:fem}, to simulate the measurements, whereas the coarse mesh was used to solve the inverse problem. We use piecewise constant basis functions for the conductivity $\sigma$ and piecewise linear basis functions for the potential $u$. Further, we added $0.5\%$ relative noise:
\begin{align}
    \mathbf{U}^\delta = \mathbf{U} + \delta ~ \text{mean}(|\mathbf{U}|) \epsilon, \quad \epsilon \sim \mathcal{N}(0,\mathbf{I}),
\end{align}
with $\delta = 0.005$. This noise level was chosen to achieve a similar SNR as the KIT4 experiments \citep{hauptmann2017open}.

\subsection{The KIT dataset} 
The KIT4 dataset is an open dataset \citep{hauptmann2017open} of EIT measurements from the Kuopio Impedance Tomograph 4 (KIT4) located at the University of Eastern Finland \citep{kourunen2008suitability}. 
The measurements were taken on a circular tank of radius 14 cm with 16 electrodes of width 2.5 cm and tap water with a conductivity value 0.03 S/m filled to a height of 7 cm. Between 1 to 3 conductive (metal) and resistive (plastic) targets were placed in the tank, with a total of 22 different measurements. Adjacent current patterns with amplitude 2 mA were applied at 1 kHz. Additionally, an out-of-distribution dataset has been collected comprising a resistive foam close to the boundary $\Gamma$ with inclusions inside (13 measurements), as well as one final measurements of a pumpkin.

The KIT4 dataset lacks ground truth conductivity distributions, making it impossible to compute the evaluation scores outlined in Section~\ref{sec:evaluation_scores}, with the exception of the measurement error. However, the dataset includes photographs of the watertank with various inclusions. These images were manually annotated to create segmentation masks, categorizing regions into resistive/conductive inclusions and background. These annotated masks were used to calculate the Dice Score, allowing a quantitative comparison of the reconstruction methods.

\section{Numerical results}    
\label{sec:numerical_experiments}
In this section we evaluate the model-based and learned reconstruction methods on three distinct datasets: the held-out test set of simulated ellipses, a small simulated dataset of differently shaped inclusions and the KIT4 dataset. In particular, we evaluate the linearized reconstruction method (\textbf{Lin-Rec}, see \eqref{eq:lin_rec_one_step}), \textbf{D-Bar} (see Section \ref{sec:Dbar}, sparsity reconstruction (\textbf{L1-Sparsity}, see Section \ref{sec:sparsity}),  Gauss-Newton TV (\textbf{GN-TV}, see \eqref{eq:tv_update}), \textbf{FC-UNet} (see Section \ref{sec:fully_learned}), post-processing UNet (\textbf{Post-UNet}, see Section \ref{sec:lin_post}), \textbf{Deep D-bar} (see Section~\ref{sec:lin_post}), deep direct sampling method (\textbf{Deep DSM}, see Section \ref{sec:deepdsm}) and the deep Gauss-Newton TV (\textbf{Deep GN-TV}, see Section \ref{sec:unrolled}). Further, we provide visual comparisons to the level set method, cf. Section \ref{sec:level_set}, for which we do not present quantitative results. All learned reconstruction methods were trained on the training dataset described in Section~\ref{sec:sim_data}. For the model-based approaches, we tuned hyperparameters, e.g., the regularization parameter, on the validation set. 

\begin{table}[t]
\centering
\caption{Performance comparison of model-based and learned methods for the full set of current patterns on the test set of the simulated ellipses. We also compare against a naive method which only outputs a constant conductivity equal to the background. We state the mean and standard deviation for all scores, except for the measurement error, where we provide the median and the $25\%$ and $75\%$ percentiles.} \vspace{0.2cm}
\resizebox{\textwidth}{!}{
\begin{tabular}{ccccccc}
\toprule
 \textbf{Method}    & \textbf{Rel.~\( L^1 \)-Error} $(\downarrow)$ & \textbf{Rel.~\( L^2 \)-Error} $(\downarrow)$ & \textbf{DS} $(\uparrow)$ & \textbf{DR} $(\approx 1)$ & \textbf{Meas.~Error} $(\downarrow)$ \\ \midrule
  \textbf{Lin-Rec}  & $0.209$ {\footnotesize $\pm 0.09$} & $0.096$ {\footnotesize $\pm 0.07$} & $0.624$ {\footnotesize $\pm 0.12$}& $0.970$ {\footnotesize $\pm 0.42$} & $0.038$ {\footnotesize $(0.001, 0.328)$} \\
  \textbf{D-bar}  & $0.320$ {\footnotesize $\pm 0.10$} & $0.145$ {\footnotesize $\pm 0.10$} & $0.514$ {\footnotesize $\pm 0.18$} & $1.130$ {\footnotesize $\pm 0.39$} & $0.116$ {\footnotesize $(0.087, 0.182)$}  \\ 
  \textbf{L1-Sparsity} & $0.109$ {\footnotesize $\pm 0.05$} & $0.057$ {\footnotesize $\pm 0.03$}& $0.788$ {\footnotesize $\pm 0.12$}  & $0.912$ {\footnotesize $\pm 0.29$}  & $7.91\mathrm{e}{-4}$ {\footnotesize $(5.48\mathrm{e}{-4},1.18\mathrm{e}{-3})$}  \\
 \textbf{GN-TV}  & $0.096$ {\footnotesize $\pm 0.03$} & $0.024$ {\footnotesize $\pm 0.01$} & $0.876$ {\footnotesize $\pm 0.08$} & $1.110$ {\footnotesize $\pm 0.19$} & \best{$1.98\mathrm{e}{-5}$ {\footnotesize $(1.75\mathrm{e}{-5},2.54\mathrm{e}{-5})$}} \\\midrule
\textbf{FC-UNet}  & \best{$0.035$ {\footnotesize $\pm 0.02$}} & \best{$0.011$ {\footnotesize $\pm 0.01$}} & \best{$0.958$ {\footnotesize $\pm 0.02$}} & $1.041$ {\footnotesize $\pm 0.09$} & $3.25\mathrm{e}{-4}$ {\footnotesize $(2.40\mathrm{e}{-4},5.14\mathrm{e}{-4})$}  \\
\textbf{Post-UNet} & $0.036$ {\footnotesize $\pm 0.02$} & $0.013$ {\footnotesize $\pm 0.01$} & $0.956$ {\footnotesize $\pm 0.02$} & \best{$0.994$ {\footnotesize $\pm 0.07$}} & $2.99\mathrm{e}{-4}$ {\footnotesize $(2.15\mathrm{e}{-4}, 6.56\mathrm{e}{-4})$}  \\
\textbf{Deep D-bar} & $0.129$ {\footnotesize $\pm 0.06$} & $0.065$ {\footnotesize $\pm 0.04$} & $0.742$ {\footnotesize $\pm 0.12$}  & $1.179$ {\footnotesize $\pm 0.34$}& $4.72\mathrm{e}{-3}$ {\footnotesize $(2.82\mathrm{e}{-3},8.16\mathrm{e}{-3})$}  \\ 
\textbf{Deep DSM}  & $0.084$ {\footnotesize $\pm 0.03$} & $0.035$ {\footnotesize $\pm 0.02$} & $0.882$ {\footnotesize $\pm 0.06$} & $1.058$ {\footnotesize $\pm 0.18$} &  $7.36\mathrm{e}{-4}$ {\footnotesize $(4.21\mathrm{e}{-4},2.19\mathrm{e}{-3})$} \\
\textbf{Deep GN-TV}  & $0.042$ {\footnotesize $\pm 0.02$} & $0.016$ {\footnotesize $\pm 0.01$}  & $0.945$ {\footnotesize $\pm 0.02$} & $1.059$ {\footnotesize $\pm 0.13$} &  $3.55\mathrm{e}{-4}$ {\footnotesize $(2.40\mathrm{e}{-4},6.61\mathrm{e}{-4})$} \\
\midrule
Const. Background  & $0.171$ {\footnotesize $\pm 0.09$} & $0.140$ {\footnotesize $\pm 0.09$} & $0.399$ {\footnotesize $\pm 0.08$} & $0.0$ {\footnotesize $\pm 0.0$} & $0.022$ {\footnotesize $(0.008, 0.494)$}  \\
\bottomrule
\end{tabular}}
\label{tab:ellipses_results}
\end{table}

\subsection{Simulated dataset}
\begin{figure}
    \centering
    \includegraphics[width=1.0\linewidth]{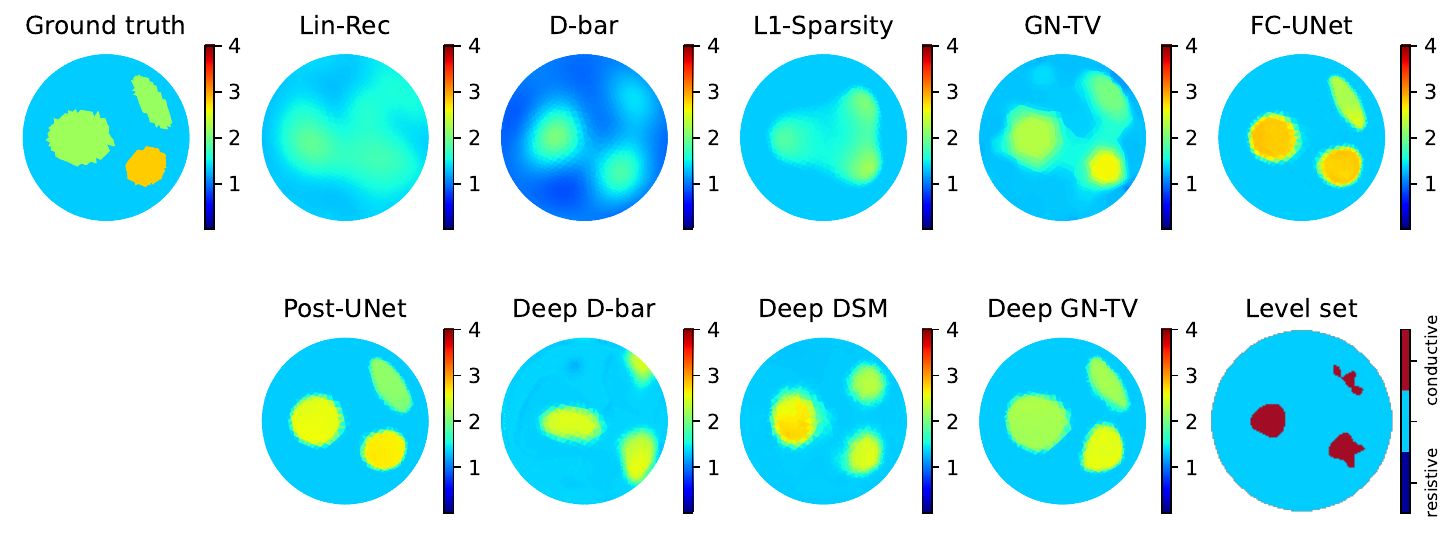}
    \rule{0.8\linewidth}{0.4pt}
    \includegraphics[width=1.0\linewidth]{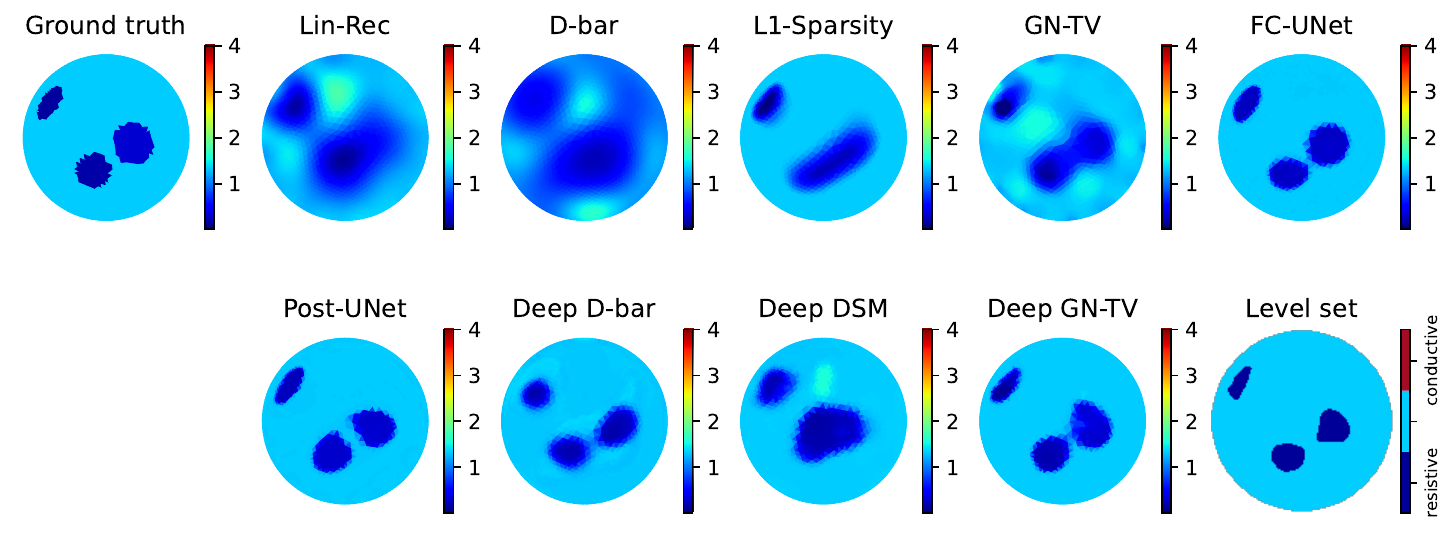}
    \rule{0.8\linewidth}{0.4pt}
    \includegraphics[width=1.0\linewidth]{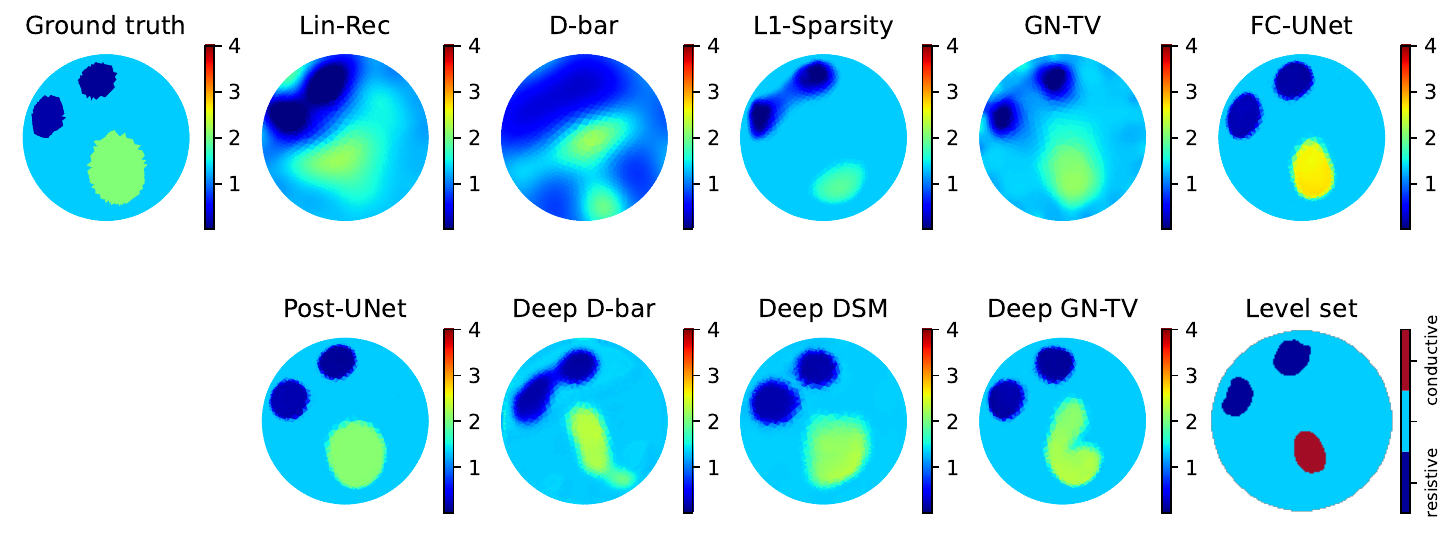}
    \caption{Reconstructions for the different methods on the test set of simulated ellipses. The level set method produce segmentations, with blue regions representing resistive inclusions and red regions representing conductive inclusions.}
    \label{fig:ellipses_reco_test}
\end{figure}

We evaluate all methods on the held-out test set of simulated ellipses. The test set shares the same characteristics, e.g., the number of inclusions, noise levels, and intensity values, as the training and validation sets, which makes it an in-distribution benchmark. Quantitative results are presented in Table~\ref{tab:ellipses_results}, including the mean and standard deviation of the evaluation scores. We observe that the measurement error is not symmetric, therefore we provide the median value and the $25\%$ and $75\%$ percentiles. Table~\ref{tab:ellipses_results} also includes the performance scores for a constant background reconstruction, which serves solely as a baseline for comparison. Exemplary reconstructions are shown in Figure~\ref{fig:ellipses_reco_test}. 

The linearized reconstruction method and D-bar produce overly smooth reconstructions and it is difficult to distinguish inclusions, as illustrated in the first row of Figure~\ref{fig:ellipses_reco_test}. The variational methods, L1-Sparsity and GN-TV, show significant improvement over these direct reconstruction techniques, although these iterative methods come with a higher computational cost. Specifically, GN-TV generates sharper conductivity distributions, though it still suffers from the typical staircasing artifacts \citep{BrediesKunischPock:2010}, one well known drawback of the total variation penalty. All model-based methods struggle to distinguish between closely located inclusions with similar intensity values, due to the severely ill-posed nature of the EIT inverse problem. However, in the visual results, the level set method is mostly able to provide a good location of the inclusions. However, for some reconstructions, e.g., the first row in Figure~\ref{fig:ellipses_reco_test}, small artifacts are clearly visible. 

In contrast, learned reconstruction methods effectively leverage the characteristics of the training set and better identify inclusions, resulting in superior performance metrics, cf. Table~\ref{tab:ellipses_results}. Both post-processing methods (Post-UNet, which takes Lin-Rec as input, and the Deep D-bar method, which uses D-bar reconstructions as input) demonstrate impressive improved performance over their model-based counterparts. In particular, the Post-UNet is able to improve the relative $L^1$-error to $0.036$, compared to $0.209$ for Lin-Rec. Similarly, Deep D-bar achieves a relative $L^1$-error of $0.129$ compared to $0.320$ for the original D-bar method. These post-processing methods are able to transform the smooth initial reconstructions into sharper outputs. (Generally better initial reconstructions lead to more accurate postprocessing results.) However, the performance is also limited by the accuracy of the initial reconstruction. The Deep DSM provides better reconstructions than Deep D-bar but falls short of the performance achieved by Deep GN-TV or FC-UNet. Additionally, while the Deep DSM reconstructions are relatively sharp, they occasionally exhibit slight distortions and fail to localize inclusions precisely (see the second row in Figure~\ref{fig:ellipses_reco_test}). The comparison between GN-TV and its learned counterpart Deep GN-TV indicates that the learned method is also able to improve the performance of the classical Gauss-Newton method. 

\begin{figure}
    \centering
    \includegraphics[width=1.0\linewidth]{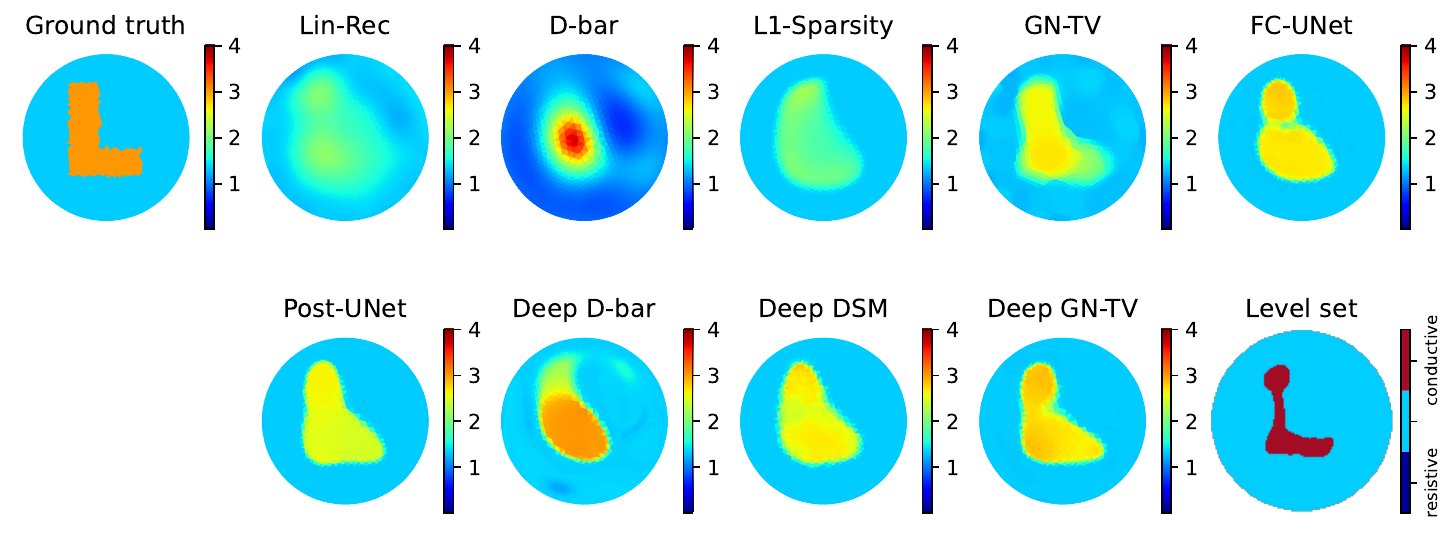}
    \caption{Reconstruction for out-of-distribution phantom. We can see that most of the learned methods stay close to the training distribution of ellipses. In particular the FC-UNet reconstructs two ellipses instead of the ''L''.}
    \label{fig:ellipses_reco_ood}
\end{figure}

The best-performing method, based on relative $L^1$ and $L^2$ errors as well as the Dice Score, is the FC-UNet on the in-distribution test set. This fully-trained method excels at recovering the locations and shapes of inclusions for phantoms with similar characteristics as the training data. However, on the out-of-distribution dataset, the performance of the FC-UNet drops significantly, see Table~\ref{tab:ellipses_results_ood}. This is attributed to the fact that the FC-UNet does not integrate any physical information of the underlying EIT inverse problem and is entirely data-dependent. In Figure~\ref{fig:ellipses_reco_ood}, one can see that the FC-UNet reconstructs two ellipses in the shape of the ''L'' instead of an actual ''L'', clearly indicating the hallucination effect on the out-of-distribution test data. Also, for this task, the level set method provides an accurate localisation, but underestimates the volume, possibly due to an imprecise choice of the penalty parameter $\alpha$.

On this generalisation task the best methods are Deep GN-TV (for relative $L^1$ error) and the Post-UNet (for relative $L^2$ error and the dice score), which directly incorporate the classical algorithms, either as part of the architecture or as the input. These observations indicate the importance of accurately incorporating the information of the physical model into the learned algorithms, in order to improve their robustness.


\begin{table}[t]
\centering
\caption{Performance comparison of model-based and learned methods for the full set of current patterns on the simulated out-of-distribution set. We also compare against a naive method which only outputs a constant conductivity equal to the background. We state the mean and standard deviation for all scores, except for the measurement error, where we provide the median and the $25\%$ and $75\%$ percentiles.} \vspace{0.2cm}
\resizebox{\textwidth}{!}{
\begin{tabular}{ccccccc}
\toprule
 \textbf{Method}    & \textbf{Rel.~\( L^1 \)-Error} $(\downarrow)$ & \textbf{Rel.~\( L^2 \)-Error} $(\downarrow)$ & \textbf{DS} $(\uparrow)$ & \textbf{DR} $(\approx 1)$ & \textbf{Meas.~Error} $(\downarrow)$ \\ \midrule
  \textbf{Lin-Rec}  & $0.168$ {\footnotesize $\pm 0.02$} & $0.076$ {\footnotesize $\pm 0.01$} & $0.581$ {\footnotesize $\pm 0.07$}& $0.586$ {\footnotesize $\pm 0.23$} & $5.56\mathrm{e}{-4}$ {\footnotesize $(3.67\mathrm{e}{-4},3.24\mathrm{e}{-3})$} \\ 
  \textbf{D-bar}   & $0.272$ {\footnotesize $\pm 0.01$} & $0.111$ {\footnotesize $\pm 0.01$} & $0.589$ {\footnotesize $\pm 0.16$}& $1.242$ {\footnotesize $\pm 0.32$} & $8.98\mathrm{e}{-2}$ {\footnotesize $(8.95\mathrm{e}{-2},9.92\mathrm{e}{-2})$}  \\ 
  \textbf{L1-Sparsity}  & $0.114$ {\footnotesize $\pm 0.05$} & $0.078$ {\footnotesize $\pm 0.03$} & $0.721$ {\footnotesize $\pm 0.17$}& $0.532$ {\footnotesize $\pm 0.27$} & $8.08\mathrm{e}{-4}$ {\footnotesize $(5.57\mathrm{e}{-4}, 9.70\mathrm{e}{-4})$}  \\
 \textbf{GN-TV}   & $0.104$ {\footnotesize $\pm 0.02$} & $0.037$ {\footnotesize $\pm 0.01$} & $0.835$ {\footnotesize $\pm 0.03$}& $0.893$ {\footnotesize $\pm 0.15$} & \best{$1.70\mathrm{e}{-5}$ {\footnotesize $(1.67\mathrm{e}{-5},1.87\mathrm{e}{-5})$}} \\\midrule
\textbf{FC-UNet}   & $0.071$ {\footnotesize $\pm 0.02$} & $0.034$ {\footnotesize $\pm 0.01$} & $0.880$ {\footnotesize $\pm 0.03$}& $0.825$ {\footnotesize $\pm 0.10$} & $2.36\mathrm{e}{-4}$ {\footnotesize $(2.22\mathrm{e}{-4},2.86\mathrm{e}{-4})$} \\
\textbf{Post-UNet} & $0.065$ {\footnotesize $\pm 0.02$} & \best{$0.031$ {\footnotesize $\pm 0.01$}} & \best{$0.891$ {\footnotesize $\pm 0.03$}} & $0.840$ {\footnotesize $\pm 0.08$} & $2.09\mathrm{e}{-4}$ {\footnotesize $(1.96\mathrm{e}{-4},2.28\mathrm{e}{-4})$} \\
\textbf{Deep D-bar}  & $0.100$ {\footnotesize $\pm 0.02$} & $0.057$ {\footnotesize $\pm 0.01$} & $0.711$ {\footnotesize $\pm 0.14$}& \best{$1.050$ {\footnotesize $\pm 0.30$}} & $7.71\mathrm{e}{-4}$ {\footnotesize $(6.18\mathrm{e}{-4}, 1.25\mathrm{e}{-3})$}  \\ 
\textbf{Deep DSM}  & $0.078$ {\footnotesize $\pm 0.02$} & $0.038$ {\footnotesize $\pm 0.01$} & $0.869$ {\footnotesize $\pm 0.03$}& $0.876$ {\footnotesize $\pm 0.12$} & $1.66\mathrm{e}{-4}$ {\footnotesize $(1.57\mathrm{e}{-4}, 1.68\mathrm{e}{-4})$} \\
\textbf{Deep GN-TV}   & \best{$0.064$ {\footnotesize $\pm 0.02$}} & $0.033$ {\footnotesize $\pm 0.01$} & $0.882$ {\footnotesize $\pm 0.03$}& $0.901$ {\footnotesize $\pm 0.10$} & $2.08\mathrm{e}{-4}$ {\footnotesize $(1.86\mathrm{e}{-4},2.45\mathrm{e}{-4})$} \\
\bottomrule
\end{tabular}}
\label{tab:ellipses_results_ood}
\end{table}

\subsection{KIT4 dataset}
We present quantitative results for the KIT4 dataset in Table~\ref{tab:kit4_results}. The KIT4 dataset is divided into subsets as follows: 'Set 1–Set 5', which include at most three different inclusions in the watertank; 'Set 6', which features a foam layer on the boundary; 'Set 7', containing a pumpkin; and 'Set 8', which has a foam layer near the boundary. The first subset 'Set 1–Set 5' is the most similar to the original training data, albeit with different shapes of inclusions. The other subsets exhibit significantly different characteristics. All models were applied directly to the KIT4 data without re-training or fine-tuning hyperparameters. However, we remind that the Dice score used here may not be the ideal measure to reflect image quality and should be interpreted as a secondary indicator of reconstruction quality. Visual inspection the qualitative reconstructions remains important for the KIT4 data.

\begin{figure}
    \centering
    \includegraphics[width=1.0\linewidth]{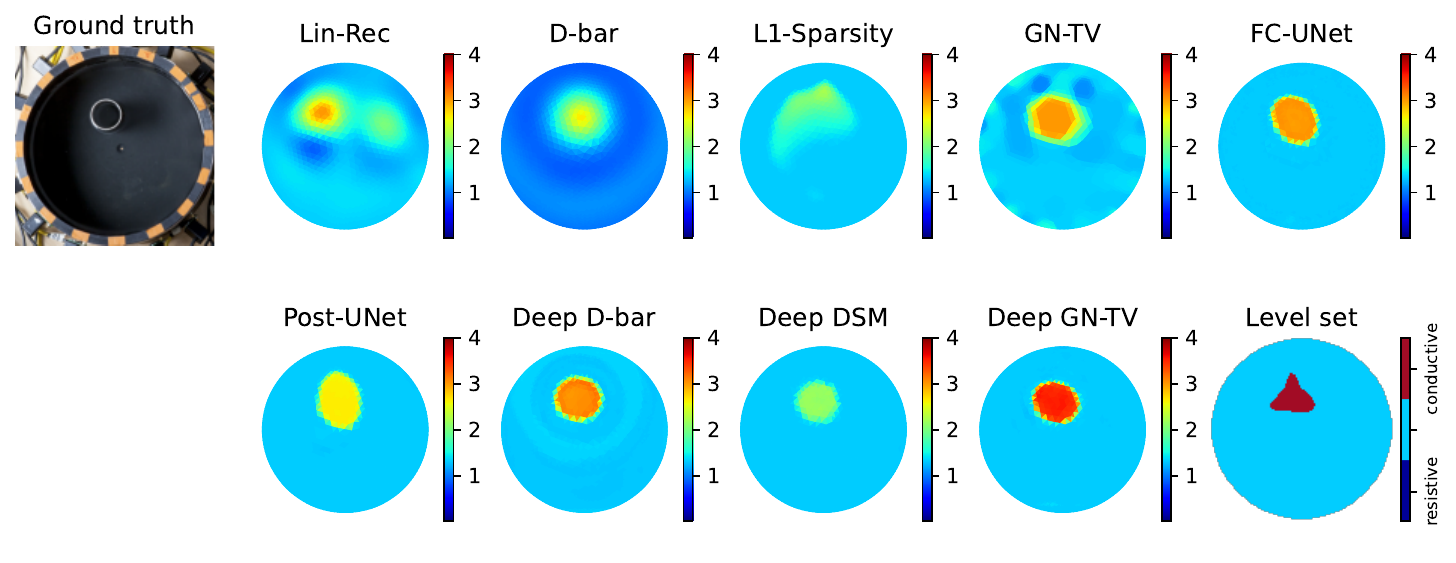}
    \rule{0.8\linewidth}{0.4pt}
    \includegraphics[width=1.0\linewidth]{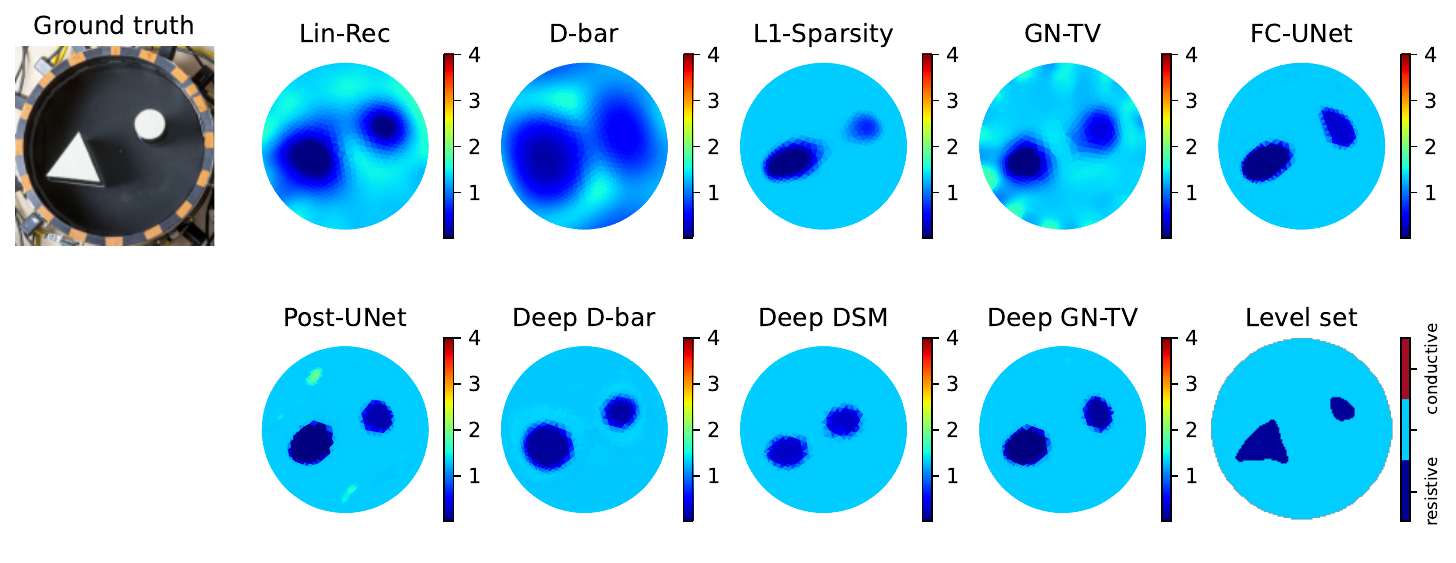}
    \rule{0.8\linewidth}{0.4pt}
    \includegraphics[width=1.0\linewidth]{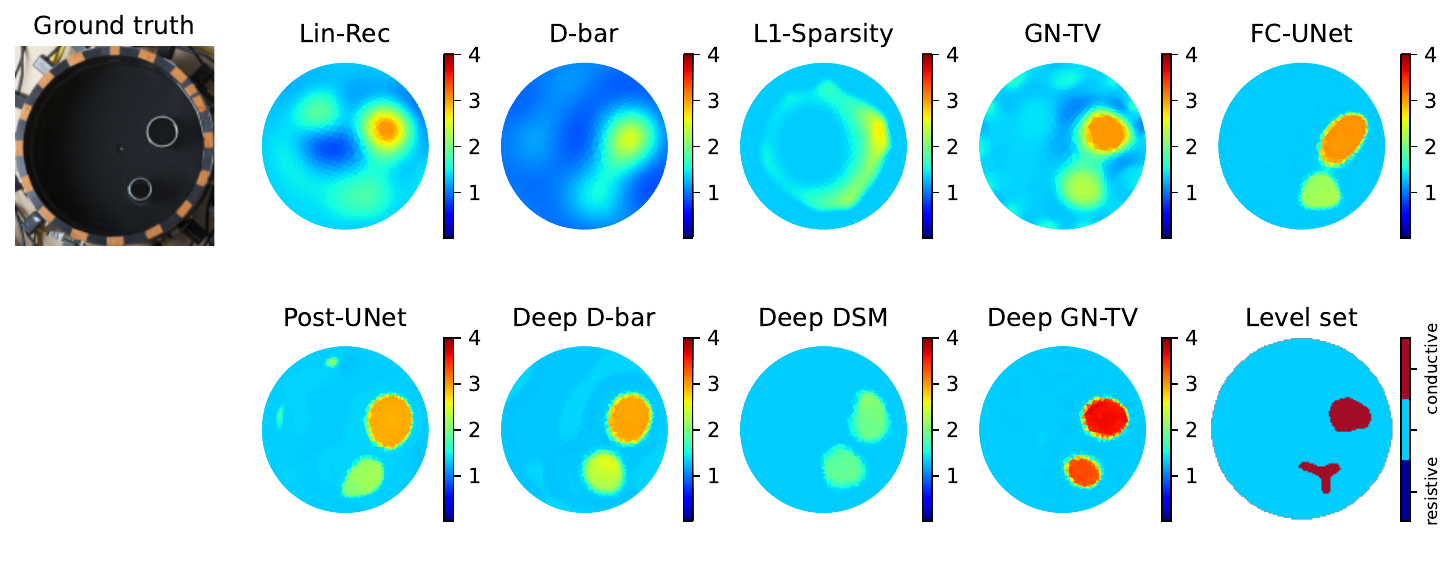}
    \caption{Reconstructions for KIT4 measurements. Top to bottom: Set 1.3, Set 2.1, Set 2.5. The level set method produce segmentations, with blue regions representing resistive inclusions and red regions representing conductive inclusions.}
    \label{fig:reco_kit4_1}
\end{figure}

\begin{figure}
    \centering
    \includegraphics[width=1.0\linewidth]{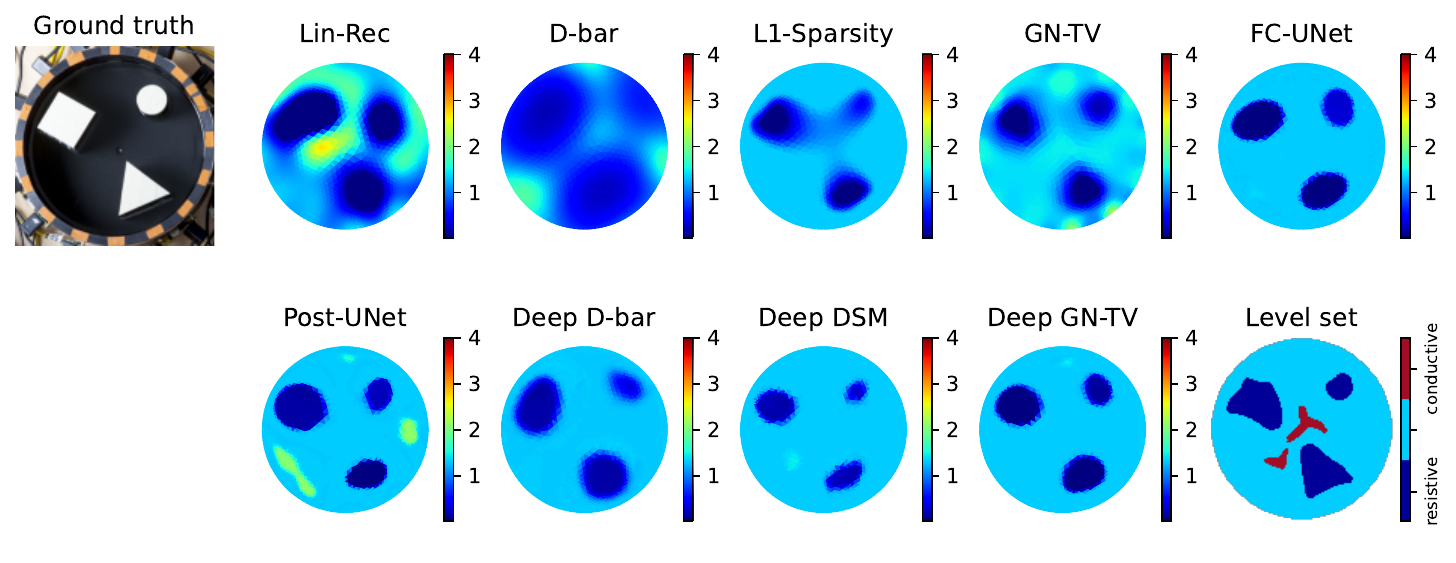}
    \rule{0.8\linewidth}{0.4pt}
    \includegraphics[width=1.0\linewidth]{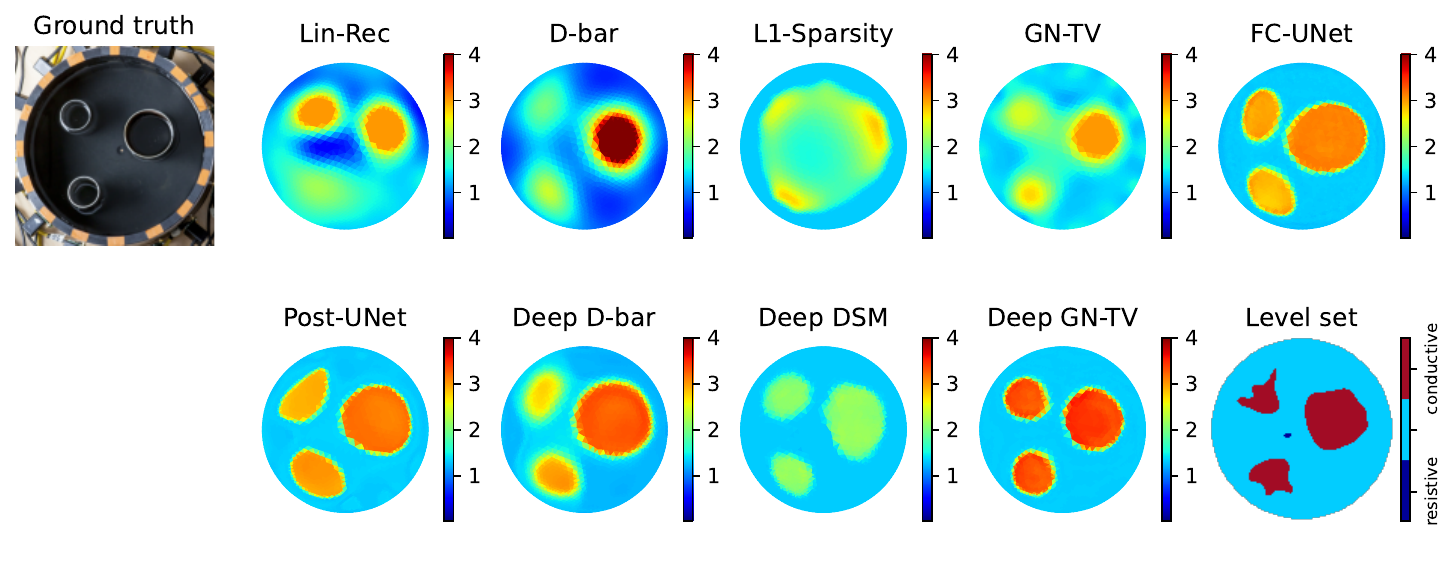}
    \rule{0.8\linewidth}{0.4pt}
    \includegraphics[width=1.0\linewidth]{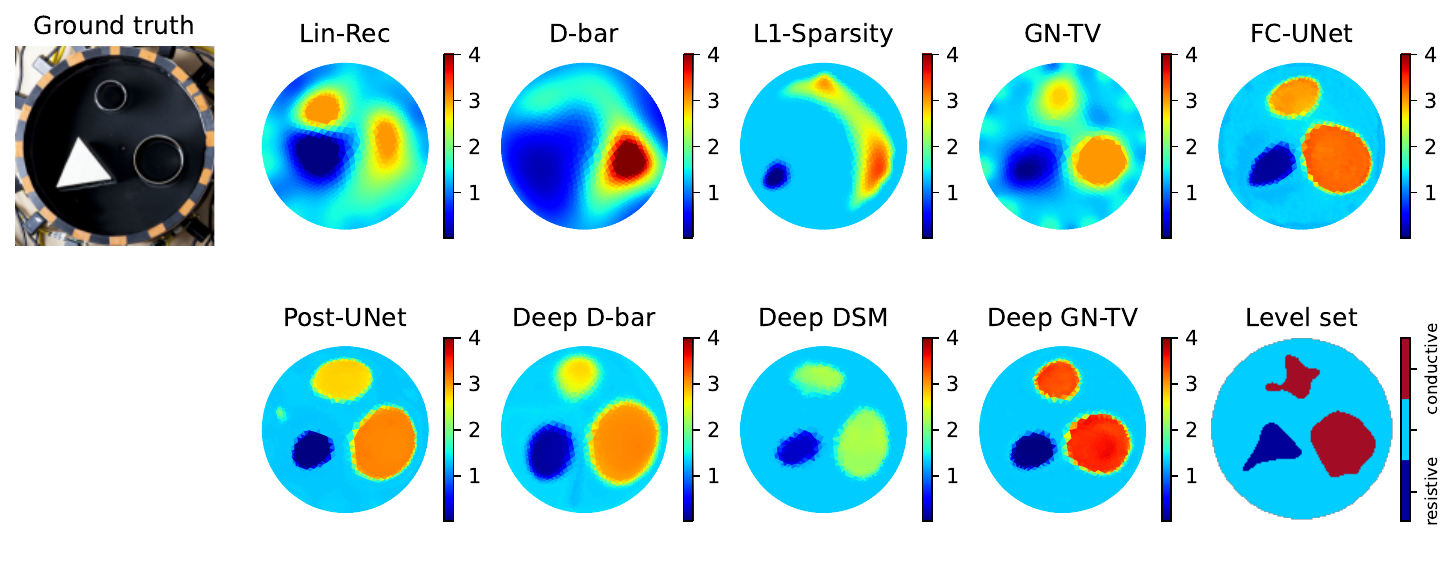}
    \caption{Reconstructions for KIT4 measurements. Top to bottom: Set 3.1, Set 3.4, Set 5.2. The level set method produce segmentations, with blue regions representing resistive inclusions and red regions representing conductive inclusions.}
    \label{fig:reco_kit4_2}
\end{figure}

For the simulated data, the FC-UNet achieves the highest performance. However, this trend does not hold for the KIT4 data. On 'Set 1–Set 5', the best-performing method is the Deep GN-TV. This is likely because Deep GN-TV incorporates classical Gauss-Newton steps, which generalize better to new data. Visual results for this scenario are presented in Figures~\ref{fig:reco_kit4_1} and \ref{fig:reco_kit4_2}. These figures show that classical methods, e.g., L1-Sparsity or GN-TV, can largely recover the inclusions but introduce a number of image artifacts. Further, for 'Set 1-Set 5' the level set method is able to produce accurate segmentations. In particular, for the second row in Figure~\ref{fig:reco_kit4_1}, the level set method is the only method which is able to recover the shape of the triangle, showing its potential for EIT.

For the remaining KIT4 subsets, all methods fail to yield satisfactory results. The presence of an additional foam layer, either on or near the boundary, drastically alters the measurement setup, and none of the methods can effectively adapt to these changes. For example, L1-Sparsity uses an $H_0^1(\Omega)$ gradient (i.e., zero-boundary conditions, and thus the conductivity distributions with inclusions on the boundary as in 'Set 8' cannot be modelled adequately, since the algorithm cannot change from the initial background conductivity. 

\begin{table}
\centering
\caption{Performance comparison of model-based and learned methods for the full set of current patterns on the KIT4 dataset. For KIT4 no ground truth data is available. We compute the Dice Score using a hand-annotated segmentation mask. Set 1 - Set 5 contains inclusions inside the water-tank. Set 6 has an additional foam layer on the boundary. Set 7 contains a slice of pumpkin. Finally, Set 8 contains additional foam near the boundary. We give the dice score (DS) and the measurement error (Meas. Error).} \vspace{0.2cm}
\resizebox{\textwidth}{!}{
\begin{tabular}{ccccccccc}
\toprule
    & \multicolumn{2}{c}{Set 1 - Set 5} & \multicolumn{2}{c}{Set 6} & \multicolumn{2}{c}{Set 7}  & \multicolumn{2}{c}{Set 8} \\ 
 \textbf{Method}    & \textbf{DS} $(\uparrow)$  & \textbf{Meas. Error} $(\downarrow)$  & \textbf{DS} $(\uparrow)$  & \textbf{Meas. Error} $(\downarrow)$ & \textbf{DS} $(\uparrow)$  & \textbf{Meas. Error} $(\downarrow)$ & \textbf{DS} $(\uparrow)$  & \textbf{Meas. Error} $(\downarrow)$ \\ \midrule
 \textbf{Lin-Rec} & $0.533$ & $0.038$ & $0.357$ & $91.99$ & $0.313$ & $0.010$ & $0.395$ & $0.173$  \\
 \textbf{D-bar}  & $0.465$ & $0.167$ & $0.251$ & $0.136$ & $0.407$ & $0.472$ & $0.425$ & $0.405$  \\ 
\textbf{L1-Sparsity}  & $0.661$ & $0.0015$ & \best{$0.603$} & $0.0031$ & $0.241$ & $0.005$ & \best{$0.629$} & $0.045$  \\
  \textbf{GN-TV} & $0.697$ & \best{$0.00037$} & $0.114$ & \best{$0.0004$} & $0.428$ & \best{$0.0002$} & $0.228$ & \best{$0.00189$}    \\
\midrule 
\textbf{FC-UNet}  & $0.761$ & $0.0014$ & $0.378$ & $0.0718$ & $0.345$ & $0.0379$ & $0.486$ & $0.496$ \\ 
\textbf{Post-UNet}  & $0.694$ & $0.001$ & $0.381$ & $0.0459$ & $0.309$ & $0.0344$ & $0.575$ & $0.0343$   \\ 
\textbf{Deep D-bar} & $0.695$ & $0.0049$ & $0.408$ & $0.0602$ & \best{$0.493$} & $0.0885$ & $0.537$ & $0.4424$   \\  
\textbf{Deep DSM} & $0.770$ & $0.0069$ & $0.350$ & $0.0419$ & $0.255$ & $0.0149$ & $0.615$ & $0.230$ \\  
\textbf{Deep GN-TV}  & \best{$0.775$}  & $0.0013$ & $0.408$ & $0.063$ & $0.276$ & $0.0124$ & $0.539$ & $0.197$  \\ 
\bottomrule
\end{tabular}}
\label{tab:kit4_results}
\end{table}

\begin{figure}
    \centering
    \includegraphics[width=1.0\linewidth]{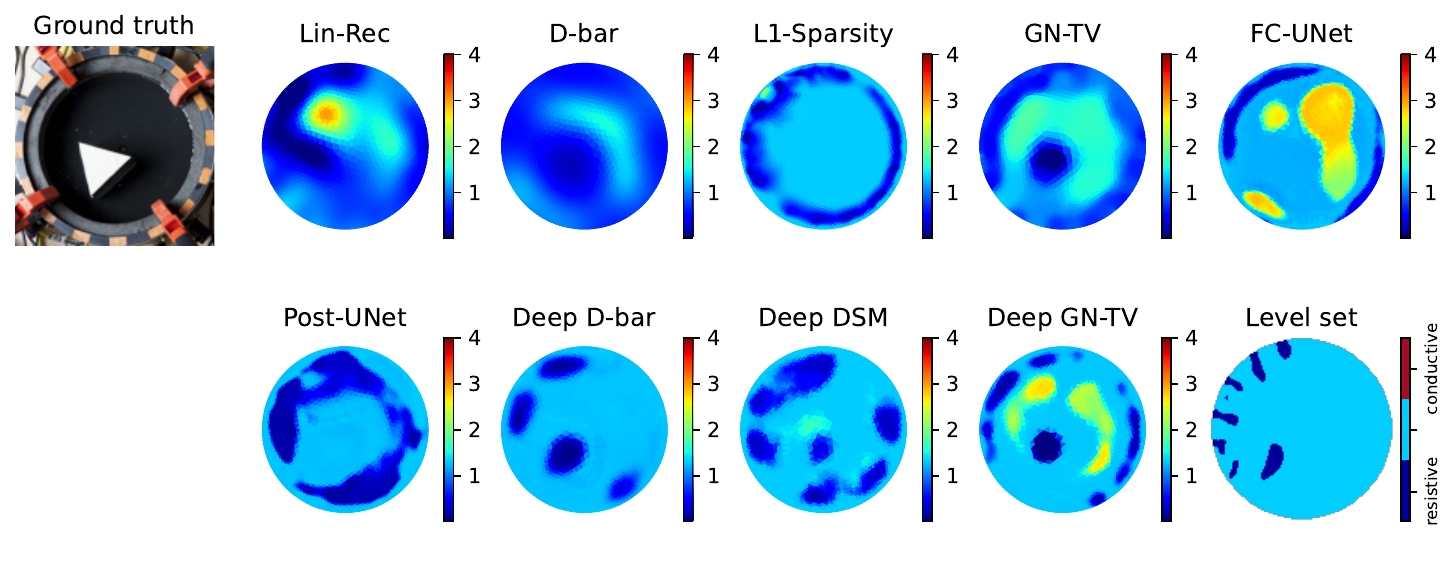}
    \rule{0.8\linewidth}{0.4pt}
    \includegraphics[width=1.0\linewidth]{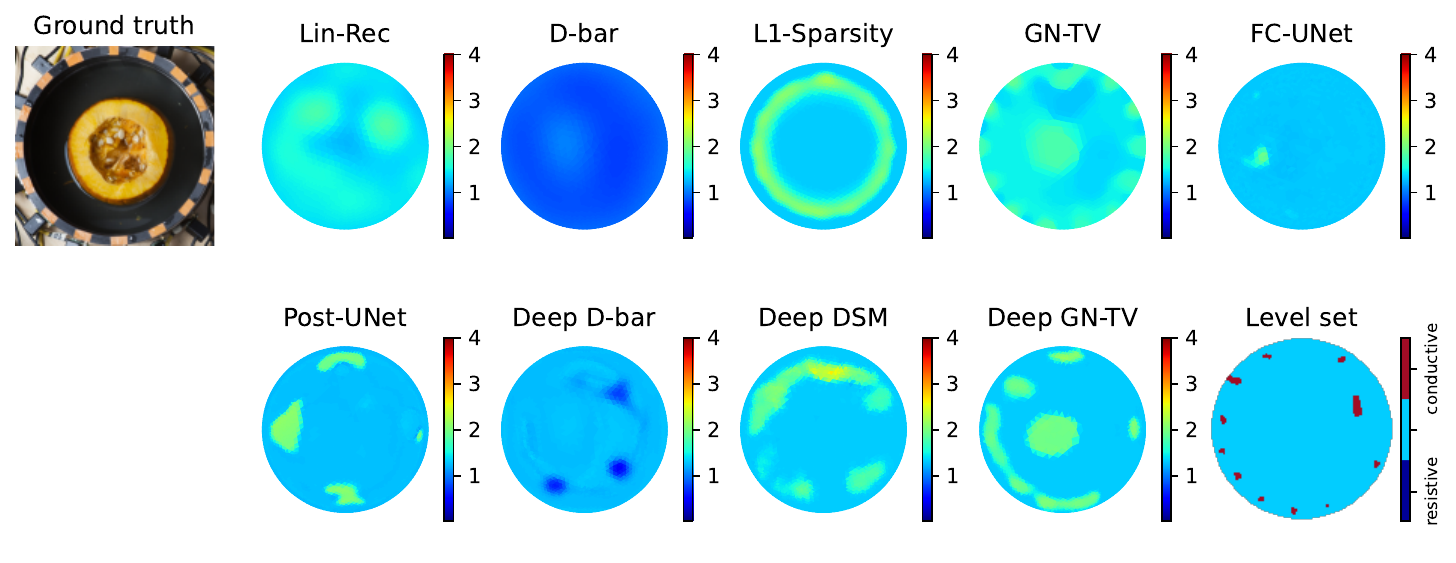}
    \rule{0.8\linewidth}{0.4pt}
    \includegraphics[width=1.0\linewidth]{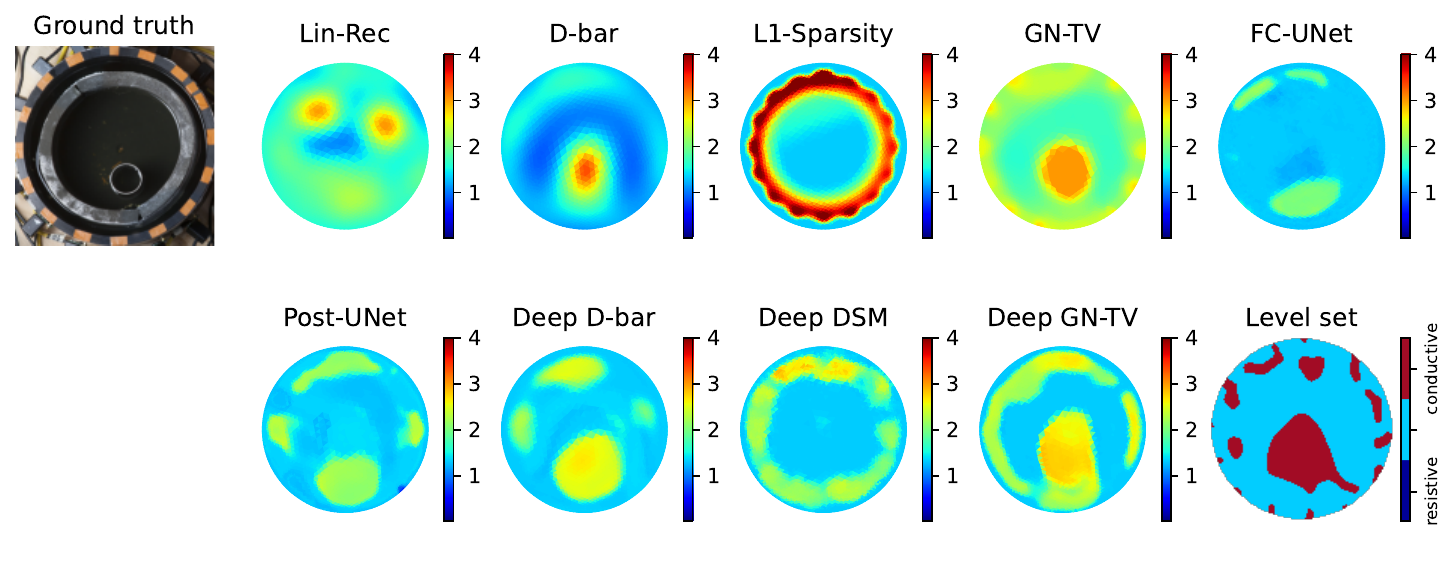}
    \caption{Reconstructions for KIT4 measurements. Top to bottom: Set 6.2, Set 7.1, Set 8.2. The level set method produce segmentations, with blue regions representing resistive inclusions and red regions representing conductive inclusions.}
    \label{fig:reco_kit4_3}
\end{figure}

\section{Conclusion} 
\label{sec:discussion}
In this paper, we have presented a numerical comparison of several widely used model-based and learned methods for image reconstruction in EIT. The study evaluates their performance across multiple criteria using three distinct datasets (both simulated and measured data), highlighting the diverse strengths of each method. The key characteristics of each approach are summarized below. It is important to note that the choice of an optimal algorithm depends heavily on the specific experimental context and the intended application of the EIT reconstruction, whether for diagnosis or further processing.

\subsection{Discussion of results on the different data sets}
Although the methods and datasets vary widely, certain general observations can be made. Let us first comment on the model-based reconstruction methods and how they perform on the standardized set of simulated ellipsoidal data. Arguably, Lin-Rec is the most commonly used model-based method for EIT and it constitutes somewhat the baseline algorithm for this application. For the given noise level of $0.5\% $ the $L^p$-error estimates are in the range of $10-20\%$ and the measurement error is approximately $4\%$, which is in the expected range for such a severely ill-posed problem and given the limited data for reconstruction. The D-bar method does not reach these values, which is primarily caused by the cut-off radius in the scattering data which leads to a strong blurring. (Nevertheless, in theory, it is a proven regularization strategy \citep{KnudsenLassasSiltanen:2009}). 
More advanced model-based reconstruction methods such as L1-Sparsity and GN-TV perform significantly better.
Not surprisingly, if GN-TV is combined with unrolling techniques, the numerical results get even better. Exploiting the training data cuts the $L^p$-errors in half for Deep GN-TV. Notably, FC-UNet and Post-UNet perform even better. Both are on equal levels overall and are the winners in terms of $L^p$-accuracy for the simulated in-distribution data.
Their error levels are between $1.5-2 \%$ for $L^2$-errors and measurement errors are below $0.1\%$.

In general, all five learned reconstruction methods based on either post-processing or unrolling surpass the model-based methods for in-distribution simulated data. However, the picture changes for out-of-distribution experiments, which test their potential for generalization. The error levels for the model-based approaches stay roughly the same, while the learned concepts show a significant increase in the error levels.
This agrees well with the intuition: the learned reconstructors were only trained on ellipsoidal data and now try to approximate the out-of-distribution examples as superpositions of ellipsoids. Still, the learned methods do slightly outperform the model-based methods. The best methods, both in terms of quantitative numerical error values and also by visual inspection, are Post-UNet and Deep GN-TV. However, we should stress that the model-based GN-TV and partially L1-Sparsity produce results in the  same range of errors, despite the differences in terms of visual inspection.

When considering the generalizability of these learned methods, it is important to take into account the number of trainable parameters. The FC-UNet has approximately 11 million parameters, while Post-UNet and Deep D-bar have about 6 million parameters each. Deep GN-TV greatly reduces the parameter count to 0.7 million. The reduced parameter count of Deep GN-TV, combined with the integration of model-based GN-TV step, likely contributes to its relatively better generalization on out-of-distribution data, as it may mitigate overfitting to the training distribution.

Both L1-Sparsity and GN-TV are very stable when they are applied to out-of-distribution data, with only a minor drop in all quality measures. This is definitely a significant advantage of model-based methods, and indeed they do not rely on previously seen similar cases and still perform very well. Meanwhile, the level set method aims at identifying the locations and shapes of the objects rather than at reconstructing the conductivity values. The  results show its general potential for EIT, and the results for the out-of-distribution example are amazing. 

Finally, let us comment on the performance on the real data set KIT4. Visual inspection of the results demonstrates the complexity of the problem. The obtained reconstructions widely vary in quality and all are open for further improvement. Specifically, the results between the simplistic targets in Figures \ref{fig:reco_kit4_1} and \ref{fig:reco_kit4_2} compared to the structured targets in Figure \ref{fig:reco_kit4_3} are drastically different. In the former case, all learned reconstruction methods performed well and generalize from the simulated training to the in-distribution test set, for the Post-UNet some residual artifacts seem to be caused by the linearized reconstructions, while Deep D-bar and Deep DSM performed best out of the post-processing type approaches. 
Notably, L1-Sparsity and Post-UNet, which did very well on simulated data, fails to resolve the inner structures of the KIT4 data set, especially for the structured targets in Figure \ref{fig:reco_kit4_3}. Similarly, the adaptation of the level set method to real data needs further refinement.

Visual inspection shows GN-TV and Deep GN-TV as clear winners: they are the only methods that allow a direct visual identification of the inner major structures. Surprisingly, this is much less mirrored by the numerical error values. Still GN-TV and Deep GN-TV give the best values, but L1-Sparsity performs very well when using dice score as error measure. This might be due to the structure of the phantoms, L1-Sparsity perfectly reconstructs the outer edges, which dominate in Dice score the inner structures. Nonetheless, in practice the full evaluation has also to be considered in view of the underlying task which usually involves further processing or classification tasks on the reconstructions.

\subsection{Summary}
Fully-learned methods, e.g. the FC-UNet, demonstrate remarkable performance on the in-distribution test set. However, one big limitation of these  methods is the strong dependence on the specific measurement settings they were trained on, and thus limited portability to different scenarios. Any change in the measurement process, e.g., varying the injection patterns and the number of electrodes, necessitates re-training the model from the scratch, which can be computationally expensive and requires a new training dataset. In contrast, hybrid approaches, e.g., Post-UNet or Deep GN-TV, offer greater adaptability. These methods can be used for different measurement settings without requiring extensive retraining, albeit sometimes at the cost of reduced reconstruction performance. This adaptability comes from the integration of the physical model (i.e., the FEM model) into the architecture, which indirectly encodes the flexibility with respect to the measurement configuration.

The ultimate goal should be to develop flexible learned reconstruction methods that can operate independently of both the measurement settings (e.g., varying injection patterns), geometry of the domain (e.g., irregular or non-circular domains) and boundary conditions (e.g., different number and size of electrodes). Achieving this level of generalization requires addressing several outstanding challenges, including designing domain independent network architectures. One option would be to make use of advances both in neural operators~\citep{azizzadenesheli2024neural} and graph neural networks to build discretization invariant architectures. 

Our experimental results highlight one notable limitation of learned methods for image reconstruction: models trained exclusively on simulated data may fail to generalize effectively to real-world measurements with new characteristics. This discrepancy show the need for methods adapting pre-trained models to new, unseen data distributions. One promising avenue to address this issue involves post-training or fine-tuning strategies, such as test-time training (TTT) \citep{darestani2022test,barbano2022educated}. These approaches allow a model to adapt dynamically, after initial training, to the distribution of the test data. For instance, leveraging an unsupervised loss function during TTT can enable fine-tuning based solely on the measured data, potentially bridging the gap between simulated and real-world data. Recent work has also shown that training DNNs in a purely unsupervised manner can be effective for EIT image reconstruction \citep{liu2023deepeit}, highlighting the potential of incorporating these unsupervised training strategies as a fine-tuning step. Future research should explore generalization capabilities, with a focus on integrating fine-tuning methods into existing learned reconstruction methods.

\section*{Acknowledgments}
Alexander Denker acknowledges support by the UK EPSRC programme grant EP/V026259/1. Bangti Jin is supported by Hong Kong RGC General Research Fund (Project 14306423 and Project 14306824), and a start-up fund from The Chinese University of Hong Kong. Andreas Hauptmann is supported by the Research Council of Finland with the Flagship of Advanced Mathematics for Sensing Imaging and Modelling Project No. 359186, Centre of Excellence of Inverse Modelling and Imaging Project No. 353093, and the Academy Research Fellow Project No. 338408. Kim Knudsen is supported by the Villum Foundation (grant no. 25893) and acknowledges computational assistance from Chao Zhang. Derick Nganyu Tanyu acknowledges support by the Deutsche Forschungsgemeinschaft (DFG, German Research Foundation) - project number 281474342/GRK2224/1 “Pi3 : Parameter Identification - Analysis, Algorithms, Applications”. Peter Maass acknowledges support by the Sino-German Center for Research Promotion (CDZ) within the Sino-German Mobility Programme "Inverse problems – theories, methods and implementations" (Grant No. M-0187). Fabio Margotti acknowledges the support of the
Brazilian funding agency National Council for Scientific and Technological Development (CNPq - Grant 406206/2021-0).

\bibliographystyle{abbrvnat}
\bibliography{references}

\end{document}